\documentclass[10pt,journal,compsoc]{IEEEtran}
\ifCLASSOPTIONcompsoc
  \usepackage[nocompress]{cite}
\else
  \usepackage{cite}
\fi
\usepackage[dvipsnames, svgnames, x11names]{xcolor}
\usepackage[numbers,sort&compress]{natbib}
\usepackage{graphicx}
\usepackage{listings}
\usepackage{amssymb}
\usepackage{listofitems}
\usepackage{booktabs} 
\usepackage{enumitem}
\usepackage{multirow}
\usepackage{eucal}
\usepackage{array}
\usepackage{makecell}
\usepackage{hyperref}
\usepackage{diagbox}
\usepackage{threeparttable}
\usepackage{pifont}
\newcommand\textlist[3][\cr]{%
  \readlist\indices{#3}%
  \foreachitem\x\in\indices{%
    \ifnum\xcnt=1\else, \fi$#2_{\x}$%
  }%
  \ifx\cr#1\relax\else%
    \if\relax#1\relax, \ldots\else, \ldots, $#2_{#1}$\fi%
  \fi%
}
\usepackage{listings, xcolor}

\definecolor{verylightgray}{rgb}{.97,.97,.97}
\definecolor{newcolor}{rgb}{88,66,155}

\lstdefinelanguage{Solidity}{
	keywords=[1]{anonymous, assembly, assert, balance, break, call, callcode, case, catch, class, constant, continue, constructor, contract, debugger, default, delegatecall, delete, do, else, emit, event, experimental, export, external, false, finally, for, function, gas, if, implements, import, in, indexed, instanceof, interface, internal, is, length, library, log0, log1, log2, log3, log4, memory, modifier, new, payable, pragma, private, protected, public, pure, push, require, return, returns, revert, selfdestruct, send, solidity, storage, struct, suicide, super, switch, then, this, throw, transfer, true, try, typeof, using, value, view, while, with, addmod, ecrecover, keccak256, mulmod, ripemd160, sha256, sha3}, 
	keywordstyle=[1]\color{blue}\bfseries,
	keywords=[2]{address, bool, byte, bytes, bytes1, bytes2, bytes3, bytes4, bytes5, bytes6, bytes7, bytes8, bytes9, bytes10, bytes11, bytes12, bytes13, bytes14, bytes15, bytes16, bytes17, bytes18, bytes19, bytes20, bytes21, bytes22, bytes23, bytes24, bytes25, bytes26, bytes27, bytes28, bytes29, bytes30, bytes31, bytes32, enum, int, int8, int16, int24, int32, int40, int48, int56, int64, int72, int80, int88, int96, int104, int112, int120, int128, int136, int144, int152, int160, int168, int176, int184, int192, int200, int208, int216, int224, int232, int240, int248, int256, mapping, string, uint, uint8, uint16, uint24, uint32, uint40, uint48, uint56, uint64, uint72, uint80, uint88, uint96, uint104, uint112, uint120, uint128, uint136, uint144, uint152, uint160, uint168, uint176, uint184, uint192, uint200, uint208, uint216, uint224, uint232, uint240, uint248, uint256, var, void, ether, finney, szabo, wei, days, hours, minutes, seconds, weeks, years},	
	keywordstyle=[2]\color{teal}\bfseries,
	keywords=[3]{block, blockhash, coinbase, difficulty, gaslimit, number, timestamp, msg, data, gas, sender, sig, value, now, tx, gasprice, origin},	
	keywordstyle=[3]\color{violet}\bfseries,
	identifierstyle=\color{black},
	sensitive=false,
	comment=[l]{//},
	morecomment=[s]{/*}{*/},
	commentstyle={\color{red!60!black}},
	stringstyle=\color{red}\ttfamily,
	morestring=[b]',
	morestring=[b]"
}

\lstdefinelanguage{SolidityNoBalance}{
	keywords=[1]{anonymous, assembly, assert, break, call, callcode, case, catch, class, constant, continue, constructor, contract, debugger, default, delegatecall, delete, do, else, emit, event, experimental, export, external, false, finally, for, function, gas, if, implements, import, in, indexed, instanceof, interface, internal, is, length, library, log0, log1, log2, log3, log4, memory, modifier, new, payable, pragma, private, protected, public, pure, push, require, return, returns, revert, selfdestruct, send, solidity, storage, struct, suicide, super, switch, then, this, throw, transfer, true, try, typeof, using, value, view, while, with, addmod, ecrecover, keccak256, mulmod, ripemd160, sha256, sha3}, 
	keywordstyle=[1]\color{blue}\bfseries,
	keywords=[2]{address, bool, byte, bytes, bytes1, bytes2, bytes3, bytes4, bytes5, bytes6, bytes7, bytes8, bytes9, bytes10, bytes11, bytes12, bytes13, bytes14, bytes15, bytes16, bytes17, bytes18, bytes19, bytes20, bytes21, bytes22, bytes23, bytes24, bytes25, bytes26, bytes27, bytes28, bytes29, bytes30, bytes31, bytes32, enum, int, int8, int16, int24, int32, int40, int48, int56, int64, int72, int80, int88, int96, int104, int112, int120, int128, int136, int144, int152, int160, int168, int176, int184, int192, int200, int208, int216, int224, int232, int240, int248, int256, mapping, string, uint, uint8, uint16, uint24, uint32, uint40, uint48, uint56, uint64, uint72, uint80, uint88, uint96, uint104, uint112, uint120, uint128, uint136, uint144, uint152, uint160, uint168, uint176, uint184, uint192, uint200, uint208, uint216, uint224, uint232, uint240, uint248, uint256, var, void, ether, finney, szabo, wei, days, hours, minutes, seconds, weeks, years},	
	keywordstyle=[2]\color{teal}\bfseries,
	keywords=[3]{block, blockhash, coinbase, difficulty, gaslimit, number, timestamp, msg, data, gas, sender, sig, value, now, tx, gasprice, origin},	
	keywordstyle=[3]\color{violet}\bfseries,
	identifierstyle=\color{black},
	sensitive=false,
	comment=[l]{//},
	morecomment=[s]{/*}{*/},
	commentstyle={\color{red!60!black}},
	stringstyle=\color{red}\ttfamily,
	morestring=[b]',
	morestring=[b]"
}

\usepackage{listings, xcolor}

\definecolor{verylightgray}{rgb}{.97,.97,.97}
\definecolor{newcolor}{rgb}{88,66,155}

\lstdefinelanguage{Python}{
	keywords=[1]{and,break,class,continue,def,yield,del,elif ,else,except,exec,finally,for,from,global,if,import,in,lambda,not,or,pass,print,raise,return,try,while,assert,with,True, False, None},
	keywordstyle=[1]\color{blue}\bfseries,
	keywords=[2]{object,type,isinstance,copy,deepcopy,zip,enumerate,reversed,list,set,len,dict,tuple,range,append,execfile,real,imag,reduce,str,repr,__init__,__add__,__mul__,__div__,__sub__,__call__,__getitem__,__setitem__,__eq__,__ne__,__nonzero__,__rmul__,__radd__,__repr__,__str__,__get__,__truediv__,__pow__,__name__,__future__,__all__},
	keywordstyle=[2]\color{teal}\bfseries,
	keywords=[3]{self},
	keywordstyle=[3]\color{violet}\bfseries,
	identifierstyle=\color{black},
	sensitive=false,
	comment=[l]{\#},
	commentstyle={\color{red!60!black}},
	stringstyle=\color{red}\ttfamily,
	morestring=[b]',
	morestring=[b]"
}

\newtheorem{definition}{Definition}
\usepackage{subfigure}
\usepackage[normal]{caption}
\usepackage{amsmath}
\usepackage{algorithmic}
\usepackage[linesnumbered,ruled,vlined]{algorithm2e}
\usepackage{url}

\usepackage{setspace}
\begin{document}

\begin{sloppypar}

\title{Demystifying Random Number in Ethereum Smart Contract: Taxonomy, Vulnerability Identification, and Attack Detection}

\author{ Peng Qian, Jianting He, Lingling Lu, Siwei Wu, Zhipeng Lu, Lei Wu, Yajin Zhou, and Qinming He 
        
\IEEEcompsocitemizethanks{

\IEEEcompsocthanksitem 
Peng Qian, Jianting He, Lingling Lu, Siwei Wu, Zhipeng Lu, Lei Wu, Yajin Zhou, and Qinming He are with the College of Computer Science and Technology, Zhejiang University, Hangzhou, 310018, China. E-mail: messi.qp711@gmail.com, jiantinghe@zju.edu.cn, lulingling@email.cufe.edu.cn, wusw1020@zju.edu.cn, alexious@zju.edu.cn, lei\_wu@zju.edu.cn, yajin\_zhou@zju.edu.cn, hqm@zju.edu.cn.
\IEEEcompsocthanksitem 
Peng Qian is also with the Collaborative Innovation Center of Artificial Intelligence by MOE and Zhejiang Provincial Government.
\IEEEcompsocthanksitem Lei Wu is also with the Key Laboratory of Blockchain and Cyberspace Governance of Zhejiang Province.
}

\thanks{Peng Qian and Jianting He are co-first authors. Yajin Zhou is the corresponding author.}

}

\markboth{IEEE Transactions on Software Engineering}
{Qian \MakeLowercase{\textit{et al.}}: Demystifying Random Number in Ethereum Smart Contract: Taxonomy, Vulnerability Identification, and Attack Detection}

\IEEEtitleabstractindextext{
\begin{abstract}
Recent years have witnessed explosive growth in blockchain smart contract applications. As smart contracts become increasingly popular and carry trillion dollars worth of digital assets, they become more of an appealing target for attackers, who have exploited vulnerabilities in smart contracts to cause catastrophic economic losses. Notwithstanding a proliferation of work that has been developed to detect an impressive list of vulnerabilities, the bad randomness vulnerability is overlooked by many existing tools. In this paper, we make the first attempt to provide a systematic analysis of random numbers in Ethereum smart contracts, by investigating the principles behind {pseudo-random number generation} and organizing them into a taxonomy. We also lucubrate various attacks against bad random numbers and group them into four categories. Furthermore, we present \emph{RNVulDet} -- a tool that incorporates taint analysis techniques to automatically identify bad randomness vulnerabilities and detect corresponding attack transactions. To extensively verify the effectiveness of \emph{RNVulDet}, we construct three new datasets: i) 34 well-known contracts that are reported to possess bad randomness vulnerabilities, ii) 214 popular contracts that have been rigorously audited before launch and are regarded as free of bad randomness vulnerabilities, and iii) a dataset consisting of 47,668 smart contracts and 49,951 suspicious transactions. {We compare \emph{RNVulDet} with three state-of-the-art smart contract vulnerability detectors, and our tool significantly outperforms them.} Meanwhile, \emph{RNVulDet} spends 2.98s per contract on average, in most cases orders-of-magnitude faster than other tools. \emph{RNVulDet} successfully reveals 44,264 attack transactions. Our implementation and datasets are released, hoping to inspire others.
\end{abstract}

}

\maketitle

\IEEEdisplaynontitleabstractindextext
\IEEEpeerreviewmaketitle

\section{Introduction}
\label{sec:introduction}

\IEEEPARstart{E}{thereum} has become one of the most prevalent blockchain platforms in recent years~\cite{ferretti2020ethereum,zhao2021temporal}, which is the first to enable the functionality of smart contracts. Ethereum smart contracts are runnable code encoding predefined rules and executing on the blockchain via the Ethereum Virtual Machine (EVM). Smart contracts make the automatic execution of contract terms possible, giving rise to a plethora of interesting decentralized applications (DApps), such as gambling game~\cite{min2019blockchain}, lottery~\cite{jia2019delottery}, decentralized finance~\cite{qian2019digital}, and distributed exchange~\cite{soni2021blockchain}.

Nowadays, smart contracts carry trillion dollars worth of digital assets, attracting numerous attacks~\cite{chen2021defectchecker,zhuang2020smart,qian2020towards}. Indeed, Ethereum has already encountered several devastating attacks on vulnerable smart contracts~\cite{perez2021smart}. The most prominent ones were the \emph{DAO} attack in 2016~\cite{dhillon2017dao} and the \emph{Parity Multisig Wallet} attack in 2017~\cite{Parity}, together causing a loss of over \$400 million at that time. In  2018, the well-known game contract \emph{Fomo3D} was compromised by hackers, who exploited the bad randomness vulnerability in the contract, leading to a loss of \$2 million~\cite{fomo3d}. Recently, cybercriminals hacked into the \emph{Cream.Finance} contract, exploiting the reentrancy vulnerability to steal more than \$130 million worth of digital assets~\cite{Inspex}. 

In response to these attacks, there have been surging research interests in detecting smart contract vulnerabilities~\cite{oyente,feist2019slither,nguyen2020sfuzz,xue2022xfuzz,liu2021combining,liu2021smart}. Existing methods can be roughly funneled into two categories. One line of work leverages static program analysis techniques~\cite{kalra2018zeus,securify,so2020verismart,reis2020tezla} (\emph{e.g.,} symbolic execution and abstract interpretation) to identify smart contract vulnerabilities. Another line of work~\cite{he2019learning,choi2021smartian,torres2021confuzzius} executes program traces dynamically and exposes potential vulnerabilities by monitoring runtime behaviors during execution. While existing efforts have identified a common set of vulnerabilities, one important category of vulnerabilities, i.e., \emph{bad randomness vulnerability}~\cite{randomness}, has been largely overlooked so far. A bad randomness vulnerability is an exploitable bug that attackers can manipulate or predict the generated {pseudo-random number} to gain benefits illegitimately. For instance, when a smart contract uses public on-chain information (\emph{e.g.,} block timestamp) as the random seed to trigger critical operations (\emph{e.g.,} asset transfer), malicious miners may manipulate the on-chain information to generate a {pseudo-random number} that is beneficial to them.

\emph{Why focus on bad randomness vulnerability.}\quad  
(1) A number of popular decentralized applications (such as \emph{UniverseGalaxy}~\cite{universegalaxy}, \emph{Fomo3D}~\cite{fomo3d_app}, \emph{CryptoKitties}~\cite{cryptoKitties}, and \emph{Dice2Win}~\cite{dice2win}) utilizes pseudo-random numbers to implement specific functions. Therefore, it is extremely important for such decentralized applications to ensure that the generated pseudo-random numbers are free of bugs. 
(2) Bad randomness vulnerability in Ethereum smart contracts has caused enormous financial losses (over \$30 million)~\cite{Dice2win_analysis,fomo3d}.
(3) Upon scrutinizing current vulnerability detection approaches for smart contracts, we empirically found that most existing tools solely support detecting a subfield of the bad randomness vulnerability such as block dependency~\cite{chen2021defectchecker,nguyen2020sfuzz,kalra2018zeus,choi2021smartian}, while lacking an in-depth and comprehensive study on the perspective of bad randomness vulnerabilities. 
These facts motivate us to demystify random numbers in Ethereum smart contracts thoroughly as well as its vulnerability identification and attack detection. 

In this paper, we investigate the prevalence of random numbers in Ethereum smart contracts. To the best of our knowledge, this is the first work to provide a systematic analysis on the inner workings of various pseudo-random number generation strategies. In particular, we summarize four categories of attacks against weak random numbers. We also present concrete examples correspondingly to illustrate the attack procedure of each kind of attack. More importantly, we propose \emph{RNVulDet} – a tool that adopts taint analysis techniques to automatically identify bad randomness vulnerabilities and detect corresponding attack transactions. \emph{RNVulDet} simulates the runtime environment of the EVM and executes a smart contract through bytecode. During the simulated execution, \emph{RNVulDet} recovers the control flow execution paths from the contract bytecode, followed by the taint analysis. Once the execution reaches the taint sink check, \emph{RNVulDet} combines the predefined vulnerability patterns to determine whether the contract has a bad randomness vulnerability.

Specifically, we implement \emph{RNVulDet} by advocating four key designs: (1) \emph{Stack State Examination.} We design a stack state examination mechanism to reduce the execution of potential repeated program paths, alleviating the path explosion problem during the simulated execution. (2) \emph{Memory Segmentation.} We introduce a memory segmentation-based alias analysis paradigm to deal with {read} and {write} dependencies in Memory. (3) \emph{Storage Key-Value Pair Comparison.} We develop a key-value pair comparison technique to handle the data dependencies in Storage. (4) \emph{Transaction Replay.} Finally, we engage in a transaction replay engine that can monitor the runtime states of a smart contract and hook into its execution to reveal random number attack instances.

To validate the usefulness of \emph{RNVulDet}, {we conduct extensive experiments, comparing with two static analyzers, {i.e.,} \emph{Slither}~\cite{feist2019slither} and \emph{Mythril}~\cite{mythril}, and one dynamic tool, i.e., \emph{ConFuziuss}~\cite{torres2021confuzzius}, on three datasets.} Experimental results confirm the effectiveness of \emph{RNVulDet} in detecting bad randomness vulnerabilities and capturing corresponding attack transactions, {i.e.,} 181 bad randomness vulnerabilities and 44,264 attack transactions have been identified, outperforming existing tools by a large margin. Moreover, \emph{RNVulDet} achieves high efficiency in analyzing smart contracts, {i.e.,} spending 2.98s per contract on average, in most cases orders-of-magnitude faster than other tools.

Our key contributions can be summarized as follows:
\begin{itemize}
\item To our knowledge, this is the first work to present a systematic analysis of random numbers in Ethereum smart contracts. Particularly, we identify common techniques used by pseudo-random number generation and organize them into a taxonomy. We also sum up four categories of attacks against bad random numbers. 
\item We present \emph{RNVulDet}, a tool that automatically identifies bad randomness vulnerabilities in smart contracts and detects random number attack transactions. \emph{RNVulDet} consists of four key components, \emph{i.e.,} stack state examination, memory segmentation, storage key-value pair comparison, and transaction replay.
\item Extensive experiments on three datasets demonstrate that our proposed method is indeed useful in detecting bad randomness vulnerabilities and capturing random number attacks. \emph{RNVulDet} surpasses state-of-the-art vulnerability detectors and overall provides interesting insights. To embrace the community, we have released the implementation of \emph{RNVulDet} and datasets at \url{https://github.com/Messi-Q/RNVulDet}.
\end{itemize}

\emph{Paper Organizations.}\quad The rest of the paper is organized as follows. In Section~\ref{background}, we briefly introduce the background of Ethereum, smart contract, Ethereum Virtual Machine, and taint analysis technique. Then, we present a fine-grained classification of pseudo-random number generation methods in Ethereum smart contracts and provide the taxonomy of random number attacks in Section~\ref{random_number}. In Section~\ref{approach}, we introduce the overall architecture and workflow of \emph{RNVulDet}. Section~\ref{evaluation} describes the dataset construction, and presents the evaluation results. In Sections~\ref{discussion} and~\ref{related_work}, we discuss possible future work and introduce related work. Finally, we conclude the study in Section~\ref{conclusion}.

\section{Background}
\label{background}
In this section, we outline the required background for understanding the settings of our work, including the descriptions of Ethereum, smart contract, Ethereum virtual machine, and taint analysis technique. 

\subsection{Ethereum and Smart Contract}
\label{Ethereum_and_smart_contract}
Ethereum is one of the representative blockchain platforms, renowned for its introduction of smart contract functionality~\cite{szabo1997formalizing}. On the platform, there exist two types of accounts: \emph{user} and \emph{contract}, both of which store an Ether (\emph{i.e.,} the cryptocurrency of Ethereum) balance and publicly reside on the blockchain. User accounts, also known as externally owned accounts (EOAs), are managed by private keys, the 64-character hexadecimal string. 

Contract accounts instead are controlled by code, which are known as smart contracts. A smart contract can implement arbitrary rules for manipulating digital assets, such as sending or receiving Ether. The defined rules in a smart contract will be strictly and automatically followed during execution, effectuating the `code is law' logic~\cite{liu2021combining}. Explicitly, the code describes an agreement between the contract account and any users that interact with it. For example, such a contract program could encode the rules of a gambling game. Since the smart contract is stored on the blockchain, it becomes immutable and its execution is guaranteed by the blockchain. Ethereum smart contracts are usually developed using a high-level programming language, typically \emph{Solidity}. Once deployed on the Ethereum, smart contracts will be broadcast to all bookkeeping nodes for backups.

\subsection{Ethereum Virtual Machine}
\label{Ethereum_virtual_machcine}
Ethereum defines a stack-based virtual machine dubbed Ethereum Virtual Machine (EVM) to determine the outcome of the execution of a smart contract. EVM is the runtime environment for the transaction execution in Ethereum. EVM embraces a set of exclusive instructions, including commonly used arithmetic operations and comparison operations. It also supports some blockchain-specific instructions like querying block numbers. The machine translates the contract bytecode into stack operations, where each operand either pops or pushes values to the stack, and each value has a 256-bit word size. In particular, EVM engages in three different structures to store data:
\begin{itemize}
\item \emph{Stack.}\quad EVM is a stack-based machine, and thus performs all computations in a data space termed Stack. All in-memory values are also stored in the stack. It has a maximum depth of 1,024 elements and supports a word size of 256 bits.
\item \emph{Memory.}\quad Memory is a word-addressed byte array. Whenever the code execution of a contract is finished, the memory is cleared out. Write operations to the memory can be of 8 or 256 bits, whereas read operations are limited to 256-bit words. 
\item \emph{Storage.}\quad Storage is organized as a key-value store, which is the only way for smart contracts to retain state across executions. Data in Storage will be permanently persisted on the Ethereum world state\footnote{ Ethereum world state refers to a data structure stored on the blockchain mapping addresses to account states. }. For instance, a gaming smart contract could leverage Storage to maintain the balance of each player.
\end{itemize} 

To prevent the contract code to be stuck in endless loops of execution, the concept of \emph{gas} has been introduced. It associates costs with the execution of every instruction. When issuing a transaction\footnote{A transaction is a type of message call that serves three purposes, including transferring Ether, deploying a smart contract, and invoking functions of a smart contract~\cite{wood2014ethereum}.}, the sender has to specify the gas fee that he/she is willing to pay to the miner\footnote{Smart contract transactions are executed by miners, which are a network of mutually distrusting nodes. } for the execution of the contract. The gas fee is computed by \emph{gas\_cost} $\times$ \emph{gas\_price}. Gas cost depends on the computational resource taken by the execution, while the gas price is offered by the transaction creators~\cite{chen2021defectchecker}. Gas price denotes the number of {Wei} (1 Wei = 10e-18 Ether) to be paid per unit of gas for all computation costs incurred as a result of the transaction execution. Generally speaking, the higher the gas price set by the transaction creators, the sooner the miners will pack the transaction into the block. 
To limit gas cost, the sender who initiates a transaction will set the gas limit, which determines the maximum gas cost. If the gas cost of the transaction exceeds its gas limit, the execution will fail and throw an out-of-gas error~\cite{grech2018madmax}. In the meantime, EVM will terminate the execution and roll back to the previous state of the transaction. For a more thorough background of EVM, we refer the readers to the Ethereum yellow paper~\cite{wood2014ethereum}, where the formal specification and detailed inner workings of EVM can be found.

\begin{figure}
\centering
\includegraphics[width=8.7cm]{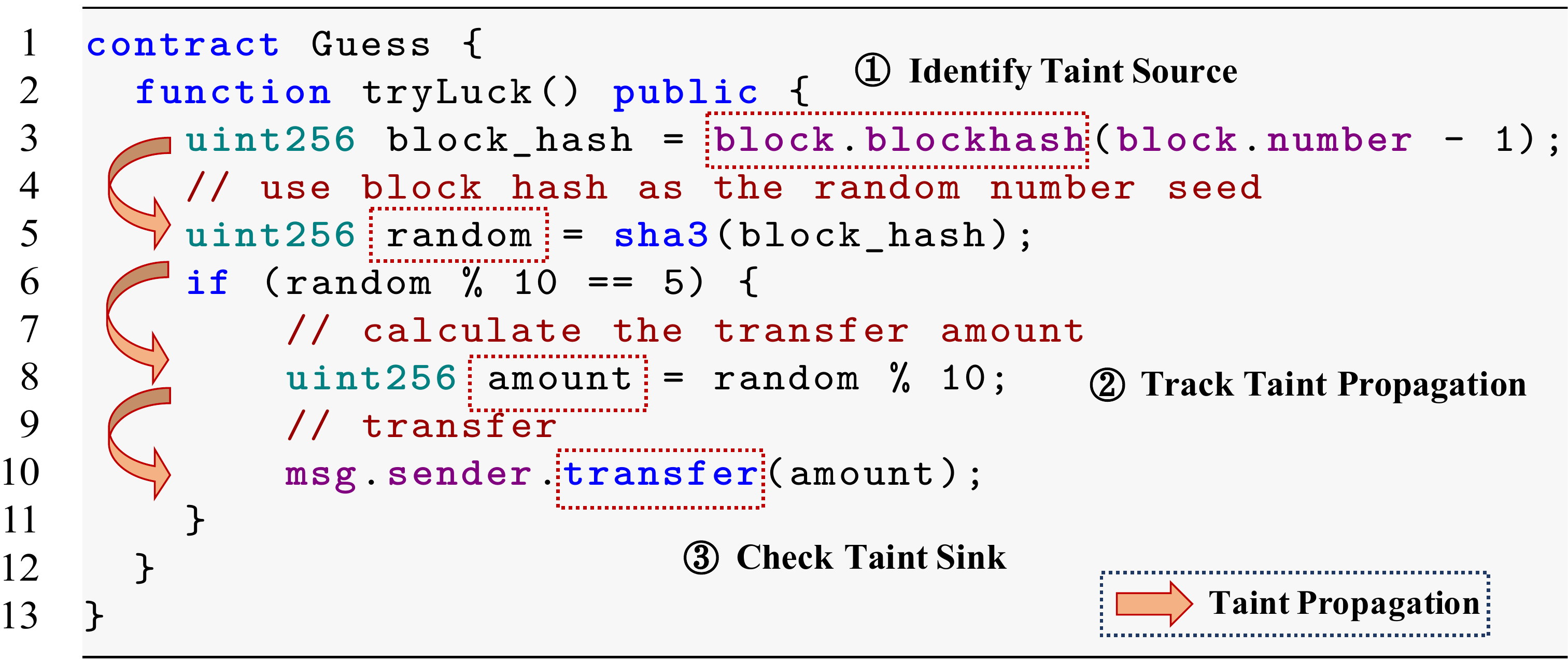}
\vspace{-0.2em}
\caption{An example of taint analysis in smart contracts.} 
\label{taint}
\vspace{-1.2em}
\end{figure}

\subsection{Taint Analysis for Vulnerability Detection}
\label{taint_analysis_technique}
Taint analysis is one of the most classical program security checking techniques~\cite{clause2007dytan,ji2021increasing}. During program execution, the taint analyzer reveals potential vulnerabilities in the code by tracking and checking the propagation trace of taint sources. Existing taint analysis techniques for program bug detection can be summarized in the following three key steps.
\begin{itemize}
\item \emph{Identify Taint Source.}\quad 
Taint sources are locations in the program where data is being read from a potentially risky source, like environment variables, command-line arguments, and file metadata. Therefore, the first key step for taint analysis is to identify such taint sources and mark them correspondingly.
\item \emph{Track Taint Propagation.}\quad
Intuitively, the marked taint sources may be assigned to other elements through data dependencies in the contract code during execution. To address this issue, the taint analyzer tracks the propagation flow of the taint sources and flags the elements affected by the taint sources accordingly. 
\item \emph{Check Taint Sink.}\quad 
From the perspective of taint analysis, the program is vulnerable if tainted elements endanger key locations, which are known as taint sinks. Every potential vulnerability has its own sinks. As such, the taint analyzer checks whether there is a trace that tainted elements can follow from taint sources to taint sinks, thereby exposing potential vulnerabilities.
\end{itemize}

Fig.~\ref{taint} presents a simplified example to illustrate the taint analysis in smart contracts. First, taint analyzer $\mathbb{A}$ identifies the taint source, i.e., \emph{block.blockhash} (line 3). Then, $\mathbb{A}$ tracks the path that the taint source travels, and marks tainted elements accordingly, i.e., \emph{random} (line 5) and \emph{amount} (line 8). Finally, $\mathbb{A}$ checks whether the \emph{transfer} operation (line 10) is the taint sink and exposes the bad randomness vulnerability. {Note that we integrate three granularities of taint propagation in the bad randomness vulnerability analysis, \emph{namely} the element-level in Stack, the memory segment-level in Memory, and the storage slot-level in Storage.}

\section{Ethereum Random Number}
\label{random_number}
In this section, we first investigate the principles behind different pseudo-random number generation strategies in Ethereum smart contracts and organize them into a taxonomy accordingly. Meanwhile, we introduce several measure indicators to evaluate each pseudo-random number generation method. Finally, we present the general process of the random number attack. We also group the existing random number attack methods into four categories and deliver concrete examples to clarify each type of attack correspondingly.

\subsection{Taxonomy of Random Number Generation}
\label{random_number_generation}

\begin{definition}[{\textbf{Pseudo-Random Number}}]\quad 
A pseudo-random number is a number generated using a set of random seeds and a mathematical algorithm that gives equal probability to all numbers occurring in the specified distribution. 
\end{definition}

After investigating existing pseudo-random number generation methods for smart contracts, we empirically found that current approaches can be roughly cast into two categories, {namely} using either on-chain information or off-chain resources to produce {pseudo-random numbers}\footnote{Resources external to the blockchain are considered as off-chain, while data stored on the blockchain is regarded as on-chain.}.

\subsubsection{Randomness Generated with On-Chain Information}
\label{on-chain}
Our first insight is to use on-chain information to generate {pseudo-random numbers} for smart contracts. Since smart contracts are programs built on the blockchain, they embrace some dedicated built-in functions to gain access to on-chain information. As the on-chain information possesses certain randomness, current contract developers often treat them as random seeds. For example, \emph{block.timestamp} which can obtain the block timestamp is usually treated as a cardinal seed to produce {pseudo-random numbers}. Typical on-chain information includes {block information, environmental information, member variables of contracts,} etc.

\renewcommand{\arraystretch}{1.12}
\begin{table*}
        \small
	\centering
	\caption{Evaluation of different random number generation strategies. ``—'' denotes not applicable.}
	\vspace{-0.9em}
	\resizebox{0.94\textwidth}{!}{
	     \begin{threeparttable}
		\begin{tabular}{cccccccccc}
			\toprule
			\textbf{Source} & \textbf{Method} & \textbf{Randomness} & \textbf{Unpredictable} & \textbf{Uncontrollable} & \textbf{Auditable} & \textbf{Open} & \textbf{Non-collusive} & \textbf{Cost} & \textbf{Latency}  \\
			\hline
			\multirow{3}{*}{{On-chain}} & Block & {\color{OliveGreen} \ding{51} }  & {\color{DarkRed} \ding{55} }    & {\color{OliveGreen} \ding{51}{*} }  & {\color{OliveGreen} \ding{51} }  & {\color{OliveGreen} \ding{51} } & — & Low & Low \\
			& Environmental & {\color{OliveGreen} \ding{51} } & {\color{DarkRed} \ding{55} }  & {\color{DarkRed} \ding{55} }    & {\color{OliveGreen} \ding{51} }  & {\color{OliveGreen} \ding{51} } & — & Low & Low  \\
			& Member Variable & {\color{OliveGreen} \ding{51} } & {\color{DarkRed} \ding{55} }  & {\color{OliveGreen} \ding{51} } & {\color{OliveGreen} \ding{51} }  & {\color{OliveGreen} \ding{51} } & — & Low & Low \\
			\midrule
			\multirow{4}{*}{{Off-chain}} & Oracle & {\color{OliveGreen} \ding{51} } & {\color{OliveGreen} \ding{51} } & {\color{OliveGreen} \ding{51}{*} }   & {\color{DarkRed} \ding{55} } & {\color{DarkRed} \ding{55} } & — & Mid & Mid \\
			& VRF & {\color{OliveGreen} \ding{51} }  & {\color{OliveGreen} \ding{51} } & {\color{OliveGreen} \ding{51} }  & {\color{OliveGreen} \ding{51} }  & {\color{DarkRed} \ding{55} } & {\color{DarkRed} \ding{55} } & Mid & Mid \\
			& Commit-Reveal & {\color{OliveGreen} \ding{51} } & {\color{OliveGreen} \ding{51} } & {\color{OliveGreen} \ding{51} }  & {\color{OliveGreen} \ding{51} }  & {\color{DarkRed} \ding{55} } & {\color{OliveGreen} \ding{51} } & Mid & Mid \\
			& VDF & {\color{OliveGreen} \ding{51} } & {\color{OliveGreen} \ding{51} } & {\color{OliveGreen} \ding{51} }   & {\color{OliveGreen} \ding{51} }  & {\color{OliveGreen} \ding{51} } & {\color{OliveGreen} \ding{51} } & High & High \\
			\bottomrule
	\end{tabular} 
	\begin{tablenotes}
	   \item {\color{OliveGreen} \ding{51}{*} }: Generally speaking, random numbers generated using block information and oracles cannot be manipulated. Exceptions exist, miners who mine the block could manipulate the block information, while oracles are able to control the random number generated by themselves.
	\end{tablenotes}
	\end{threeparttable}
	} 
\vspace{-1.2em}
\label{table1}
\end{table*}

\begin{figure}
\centering
\includegraphics[width=8.75cm]{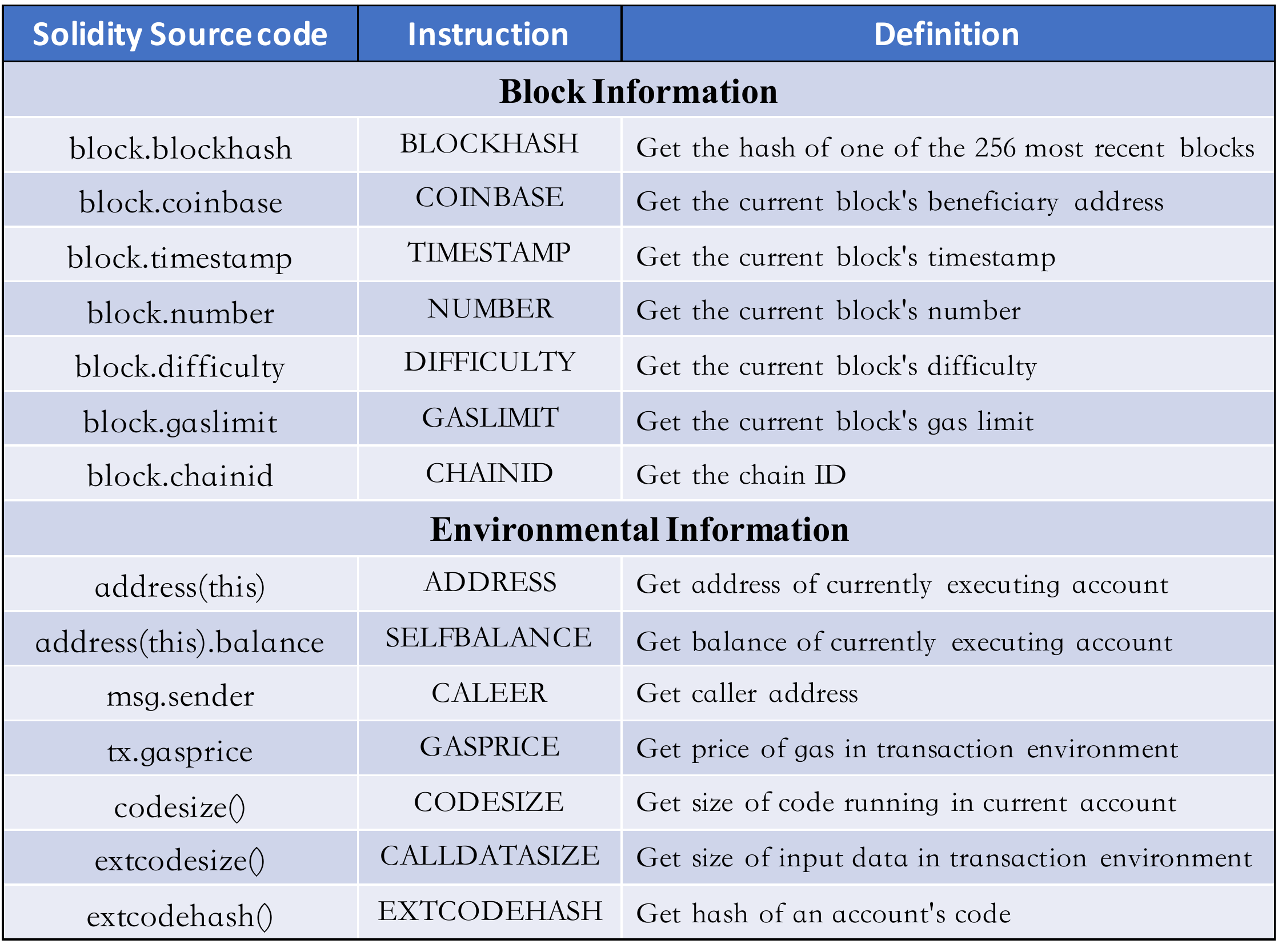}
\vspace{-0.2em}
\caption{Block information and environmental information in Ethereum, where more details can be found at~\cite{wood2014ethereum}.} 
\label{fig}
\vspace{-1.2em}
\end{figure}

\emph{Block Information.}\quad Block information refers to the properties of blocks in the blockchain, such as {block hash, block timestamp, block number}, and {block gas limit}. Fig.~\ref{fig} shows the relevant block information in Ethereum. Due to the randomness of block information, they are usually preferred as random seeds. A smart contract is able to gain such information of the {current} block where the transaction is taking place through designated instructions. However, malicious attackers can also access the block information with these instructions in an attempt to manipulate or predict the generated {pseudo-random numbers.} Another line of schemes attempts to use the block information of a future block as a basic seed, thus preventing attackers from predicting pseudo-random numbers. Unfortunately, miners who mine blocks are capable of altering the block information of the {future} block within a short time interval (roughly 900 seconds)~\cite{jiang2018contractfuzzer}. Malicious miners may thus manipulate the block information to generate a pseudo-random number that is beneficial to them.

\emph{Environmental Information.}\quad Common environmental information in Ethereum includes the address of a caller who issues the transaction, the gas price, the residual amount of gas, the size of the code running in the current environment, and so on. These kinds of environmental information have a certain degree of uncertainty, which is why they are usually regarded as random seeds to generate pseudo-random numbers. However, environmental information can be easily accessed and manipulated by attackers using simple tricks~\cite{he2020smart}. For example, an attacker can derive the address of a caller contract by using the \texttt{CREATE2} instruction~\cite{Create2}. For the residual amount of gas, an attacker can measure how much the execution is going to use beforehand, and simulate the execution locally with an instrumented EVM, thereby knowing the residual amount of gas at a given moment in time during the execution.

\emph{Contract Member Variable.}\quad In smart contracts, the types of member variables are divided into public and private. External contracts cannot directly access private member variables in the contract. As such, developers consider taking private member variables as random seeds. However, information on the blockchain is completely public. As a result, anyone can get the values of private member variables in the contract by means of off-chain access. As an example, \emph{web3.js} is a collection of libraries that allows users to interact with a local or remote Ethereum node~\cite{web3}. With function \emph{web3.eth.getStorageAt}, users have the capacity to gain access to any storage content of the contract.

\emph{Other On-Chain Information.}\quad Some developers may adopt the frequently-updating information in the contract as a random seed, like the total amount of token issuance in a token contract. Another consideration is to package the logic of pseudo-random number generation into a closed-source contract, reducing the risk of pseudo-random numbers being predicted. Nevertheless, for smart contracts, closed-source does not imply that the logic of the contract is hidden. Attackers are able to access the bytecode of the contract. Moreover, many decompilation tools, such as Erays~\cite{zhou2018erays} and Gigahorse~\cite{grech2019gigahorse}, can be used to help attackers understand the logic of the closed-source contract, so as to grasp the principles behind the generated pseudo-random numbers.

\subsubsection{Randomness Generated with Off-Chain Resources}
\label{off-chain}
Our second insight is that smart contracts generate pseudo-random numbers with the assistance of off-chain resources. Considering that adversaries can easily manipulate on-chain information, developers resort to off-chain resources for safer pseudo-random number generation. After a detailed investigation, we found that using off-chain resources for pseudo-random number generation can be summarized into four types, such as oracle, verifiable random function, Commit-Reveal, and verifiable delay function. In what follows, we will elaborate on the details of these methods.

\emph{Oracle.}\quad {Oracles are entities that connect the blockchain to external systems, allowing smart contracts to get access to off-chain resources like asset prices, token exchange rates, and random number seeds. For example, Provable~\cite{provable}, which serves as a popular oracle, provides random seeds to smart contracts by requesting external random number generation websites. Unfortunately, using oracles to generate pseudo-random numbers thus suffers from the inherent problem of whether the data obtained from oracles is trustworthy. An oracle with deprecated or malicious content could result in disastrous impacts on smart contract applications that employ the oracle information to generate pseudo-random numbers. For instance, a synthetic asset issuance platform \texttt{Synthetix} experienced an oracle attack in 2019~\cite{synthetix}, which leads to an incorrect conversion rate for sETH tokens, netting the attacker over 37 million sETH. In 2020, the decentralized smart contract application \texttt{bZx} was compromised by attackers~\cite{bzx}, who exploited the price oracle manipulation to steal 2,378 ETH. These two cases are not isolated and many smart contracts fell victim to the oracle manipulation attack, including \texttt{yEarn}~\cite{yVault}, \texttt{Harvest}~\cite{HARVEST}, \texttt{Uniswap}~\cite{Uniswap}, etc. As countermeasures, recent works such as \textsc{BlockEye}~\cite{wang2021blockeye} and \textsc{ProMutator}~\cite{wang2021promutator} have explored detecting the oracle manipulation attack. While oracle attack detection techniques have been constantly investigated, it is still in its infancy and leaves a large room for progress.}

\emph{Verifiable Random Function.}\quad Verifiable random function (VRF) is a public-key pseudo-random function that was initially put forth by Micali \emph{et al.}~\cite{micali1999verifiable}. Since then, VRF has witnessed widespread use in practice, such as lottery, cryptocurrency, and blockchain consensus. Given any input as the random seed, VRF will map the seed to a verifiable random number which is encoded with a private key signature. Public verifiers have the capacity to validate the authenticity and legitimacy of the pseudo-random number via the public key. As such, the pseudo-random number generated by VRF is unpredictable. Unfortunately, there exists the risk of collusion between the one who provides the seed input and the other who performs the signature, which threatens the security of generated verifiable random numbers. Chainlink~\cite{breidenbach2021chainlink}, a well-known decentralized oracle, leverages verifiable random functions to generate cryptographically secure randomness on-chain, which can be used by smart contract developers as a source of the tamper-proof random-number generator.

\emph{Commit-Reveal.}\quad Commit-Reveal~\cite{zhang2022f3b} is a multi-party random number generation scheme, which consists of two phases, i.e., \emph{commit} and \emph{reveal}. In the commit phase, every participant generates a random seed and calculates the corresponding hash value. All parties are required to send hash values to a proxy contract for safekeeping. In the reveal phase, each party hands over the generated random number seed to the proxy contract, which will verify whether the hash value of the seed matches the hash value submitted at the commit phase. Thereafter, random number seeds of all parties are combined to generate the final random number. 

Commit-Reveal scheme ensures the unpredictability of generated pseudo-random numbers and prevents the leakage of random seeds to a certain extent. However, due to the openness and transparency of blockchains, the participant who is the last to submit the random seed can know the seeds disclosed by other parties, so that the final random number can be calculated in advance. As such, the last participant may refuse to submit his/her random number seed once the final random number is unfavorable to him/her, hindering the generation of random numbers.

\emph{Verifiable Delay Function.}\quad Verifiable delay function (VDF) is a function whose evaluation requires running a given number of sequential steps, yet the result can be efficiently verified~\cite{wesolowski2019efficient}. VDF has been used to enhance some multi-party random number generation schemes. For example, for the Commit-Reveal, if the combined final random number is put into the VDF, and the time parameter of the VDF is set long enough (\emph{i.e.,} after the deadline of seed submission in the reveal phase), then even the last participant cannot foresee the result of the final random number. Moreover, VDF is useful for some scenarios which obtain random numbers from public sources. However, VDF is still in its infancy stage, enduring high overhead and latency.

\subsection{Evaluating Random Number Generation Method}
\label{evaluation_rand}
In this subsection, we present the evaluation results of different pseudo-random number generation strategies in Ethereum smart contracts. Specifically, we screen out 50 popular smart contracts such as \emph{UniverseGalaxy}~\cite{universegalaxy}, \emph{CryptoKitties}~\cite{cryptoKitties}, and \emph{Dice2Win}~\cite{dice2win}, and manually analyze the pseudo-random number generation methods adopted by these contracts. Quantitatively, $80\%$ of smart contracts create pseudo-random numbers by using on-chain information, of which over $70\%$ rely on block information, $8\%$ utilize environmental information, and $18\%$ focus on private member variables. On the other hand, the proportion of using off-chain resources to produce pseudo-random numbers accounts for $18\%$, of which $6\%$ request off-chain oracles and $12\%$ adopt the Commit-Reveal scheme to generate pseudo-random numbers. Note that a smart contract may employ multiple sources for producing pseudo-random numbers. As an example, contract \emph{UniverseGalaxy} generates pseudo-random numbers by combining private member variables, caller address, and previous block hash.

Further, we propose eight indicators to evaluate the aforementioned pseudo-random number generation strategies. Table~\ref{table1} shows the evaluation results of each method. From the table, we observe that using on-chain information for generating a pseudo-random number has open and auditable nature, as well as low costs and latency. However, they still suffer from inherent problems of easily being predicted and even manipulated. In contrast, the pseudo-random number generated using off-chain resources is almost unpredictable and uncontrollable but leads to higher costs and latency. Particularly, using verifiable random functions for pseudo-random number generation may endure potential collusive risks. It is worth noting that all the investigated methods possess a certain degree of randomness since every strategy integrates hash algorithms.

\subsection{Random Number vulnerability and Attack}
\label{attack_class}
\begin{definition}[\textbf{Bad Randomness Vulnerability}]\quad 
When a smart contract relies on randomness (or a pseudo-random number) to generate a value or make a decision, but the sources of the randomness are not truly random or can be compromised or predicted by bad actors, a bad randomness vulnerability may occur. 
\end{definition}
\begin{definition}[\textbf{Random Number Attack}]\quad 
A random number attack refers to the exploitation of randomness vulnerabilities to predict or manipulate the generated pseudo-random number to steal illegal profits.
\end{definition}

\begin{figure}
\centering
\includegraphics[width=8.72cm]{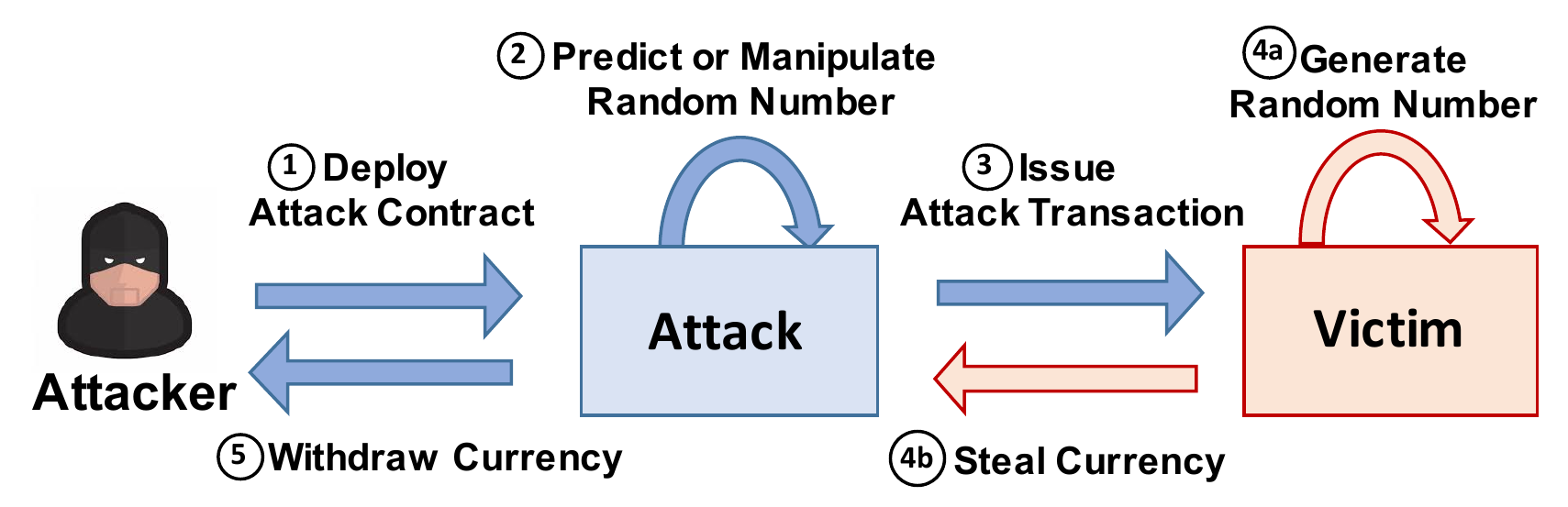}
\vspace{-0.2em}
\caption{{The general process of a random number attack.}} 
\label{attack}
\vspace{-1.2em}
\end{figure}

Upon reviewing the 50 popular smart contracts that have pseudo-random number generators, we know that most smart contracts tend to leverage on-chain information to produce pseudo-random numbers. As aforementioned, using merely on-chain information for pseudo-random number generation is insecure, which may leave a large room for security attacks. Generally, the overall process of the random number attack consists of five steps, which are depicted in Fig.~\ref{attack}.
\begin{itemize}[wide=4pt, topsep=3pt, itemsep=1.5pt, leftmargin=\dimexpr\labelwidth + 3\labelsep\relax]
\item The attacker deploys an \emph{Attack} contract on Ethereum;
\item The \emph{Attack} contract predicts or manipulates the generated pseudo-random number in a \emph{Victim} contract;
\item The \emph{Attack} contract issues an invocation (i.e., attack transaction) to the \emph{Victim} contract;
\item The \emph{Victim} contract generates a pseudo-random number and checks the invocation results of the \emph{Attack} contract. If the invocation passes, the \emph{Attack} contract steals digital currencies from the \emph{Victim} contract;
\item The attacker withdraws money from the \emph{Attack} contract.
\end{itemize}

After investigating a variety of existing random number attacks, we summarize them into four categories, viz, transaction input manipulation, random number seed manipulation, direct prediction, and transaction rollback. It is worth noting that these four classes of random number attacks cover most of the attack situations. In the following, we will deliver concrete examples to illustrate each kind of random number attack.

\begin{figure}
\centering
\includegraphics[width=8.6cm]{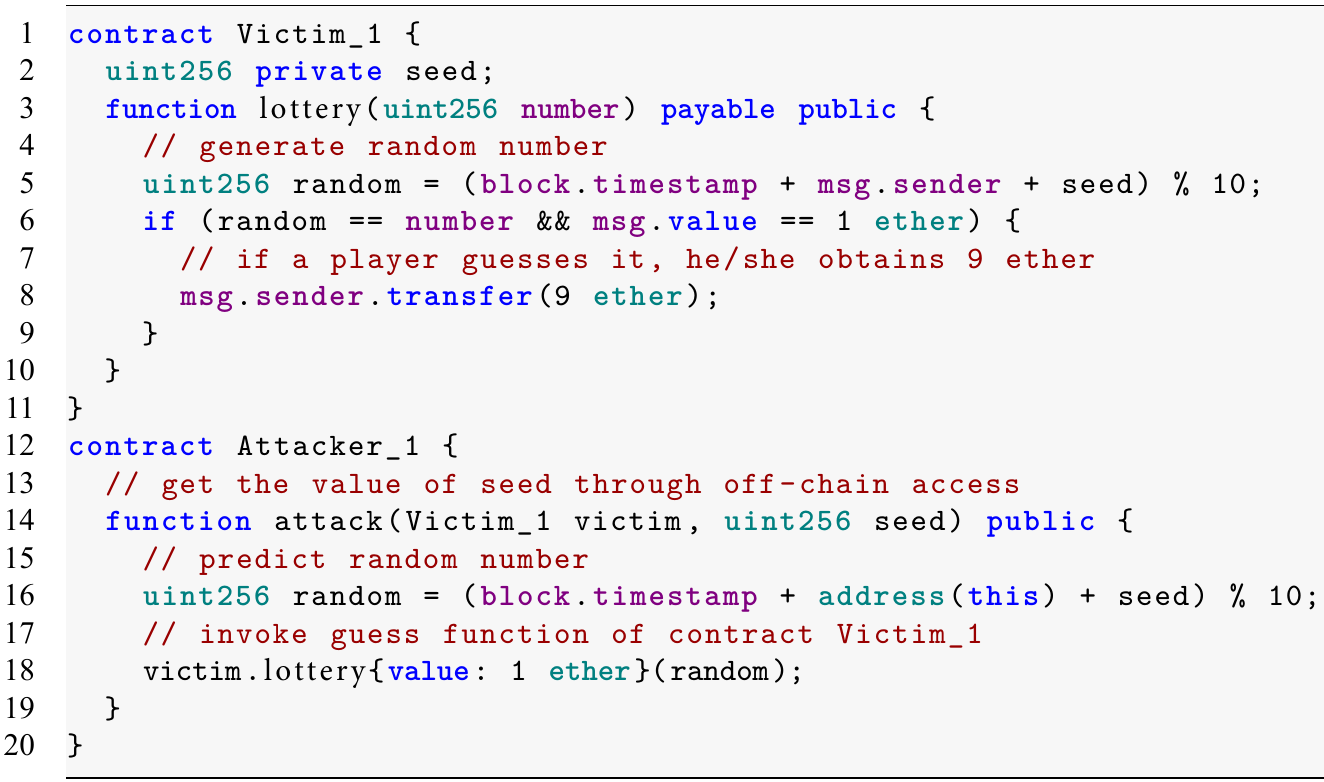}
\vspace{-0.2em}
\caption{A simplified example to show a transaction input manipulation attack.} 
\label{fig1}
\vspace{-1.2em}
\end{figure}

\subsubsection{Random Number Attack Classification}
\label{classification_attack}
In this subsection, we take the \emph{lottery} contract as an example to introduce each type of random number attack in detail.

\emph{Transaction Input Manipulation.}\quad 
Fig.~\ref{fig1} shows the simplified code of a gambling contract, which is written in Solidity. This contract realizes a lottery function that allows users to play by submitting their guesses along with participation fees. Function \emph{lottery} combines the block timestamp \emph{block.timestamp}, the caller address \emph{msg.sender}, and the private member variable \emph{seed} as the cardinal seed to generate the random number, {i.e.,} variable \emph{random}. The player who wants to participate in the game and submit a \emph{guess} needs to pay 1 ether. If the \emph{guess} number submitted by the player exactly matches the {random} number, the player will obtain 9 times the participation fee in return. 

\emph{Attack.}\quad The attacker deploys an attack contract \emph{Attacker}\_1 to predict the {random} number. Due to the openness of Ethereum blockchain, the attacker can get the value of the private member variable \emph{seed} in the victim contract \emph{Victim}\_1 through off-chain access. Then, he/she uses the value as the \emph{input} of the \emph{attack} function in \emph{Attacker}\_1 (line 14). After that, the function \emph{attack} predicts the {random} number by combining the timestamp of the current block, its contract address (i.e., \emph{address(this)}), and the variable \emph{seed} (line 16), and then invokes the function \emph{lottery} of contract \emph{Victim}\_1 (line 18). Since smart contract transactions have the atomic nature (\emph{i.e.,} all operations in a transaction must be executed successfully, otherwise, the transaction reverts to its previous state), the entire process of function \emph{attack} calling function \emph{lottery} will be completed within the same transaction. Therefore, the block timestamps obtained by functions \emph{attack} and \emph{lottery} are identical. Furthermore, for function \emph{lottery}, the address of the current contract caller (i.e., \emph{msg.sender}) is exactly equal to the address of the contract \emph{Attacker}\_1. As a result, the attacker is able to successfully predict the {random} number generated by function \emph{lottery}.

\begin{figure}
\centering
\includegraphics[width=8.55cm]{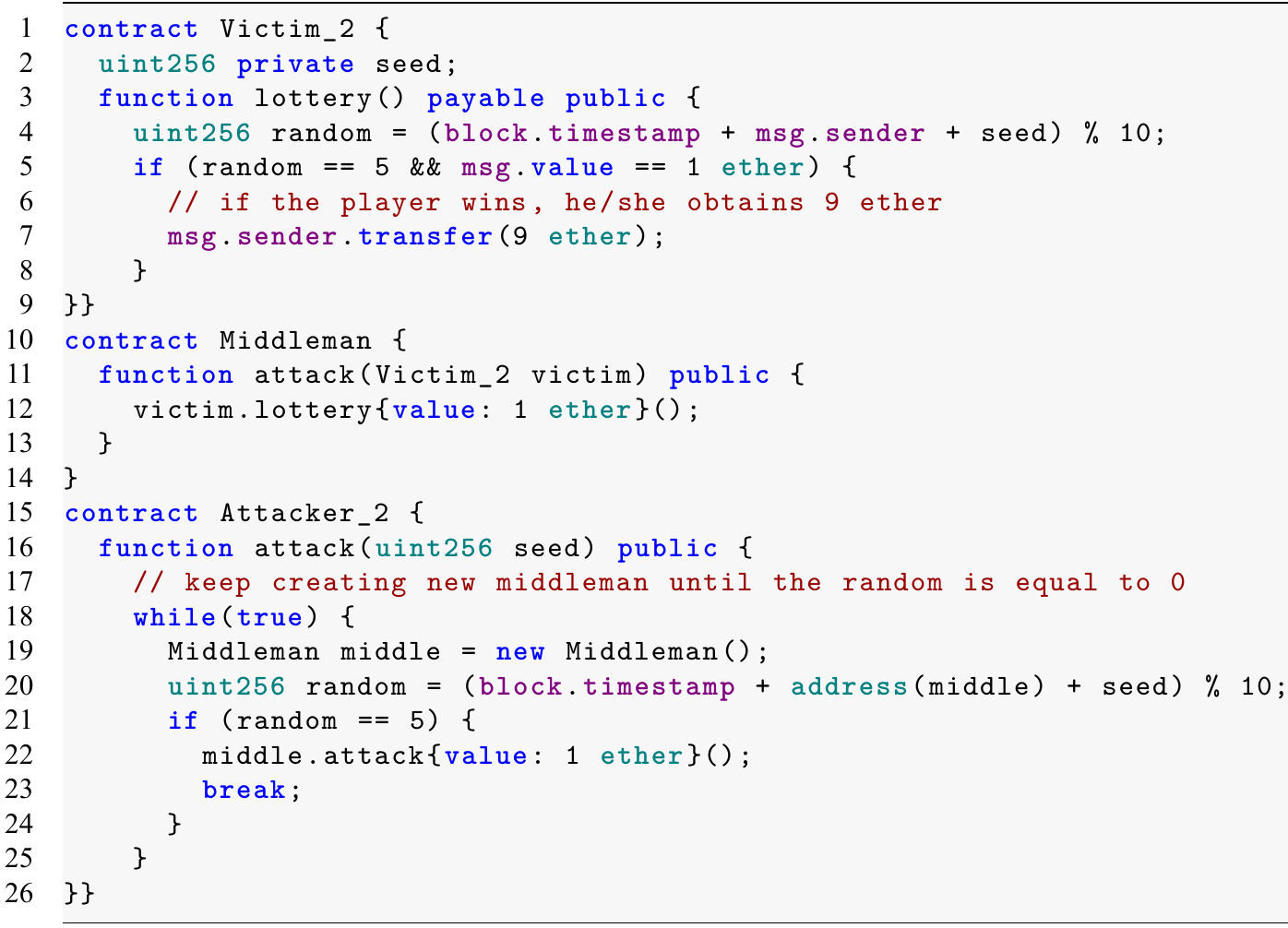}
\vspace{-0.2em}
\caption{A simplified example to show a random number seed manipulation attack. } 
\label{fig2}
\vspace{-1.2em}
\end{figure}

\emph{Random Number Seed Manipulation.}\quad 
We further present a updated gambling contract, as shown in Fig.~\ref{fig2}. The contract implements a \emph{lottery} function that users play by submitting only the participation fee of 1 ether. Function \emph{lottery} combines the block timestamp, the caller address, and the private contract member variable to generate a random number, i.e., \emph{random}. If \emph{random} is exactly equal to 5, the player who participates in the current round lottery will obtain 9 times the participation fee in return (lines 5-7).

\emph{Attack.}\quad Since function \emph{lottery} does not require any function inputs, it is impossible to predict the generated random number of function \emph{lottery} by manipulating the transaction input. However, attackers could achieve the random number attack by directly manipulating the generation of the random number seed. First, the attacker takes the value of \emph{seed} as the input (line 16). After that, the attacker creates an intermediary contract recurrently (line 19) until he/she finds an intermediate contract address so that the generated random number is equal to 5 (line 21). In this attack, attackers can easily obtain the values of \emph{block.timestamp} and \emph{seed} (which is mentioned in the previous attack). Therefore, they are only required to control the caller address (i.e., \emph{msg.sender}) to successfully predict the random number. {It is worth mentioning that the well-known \emph{Fomo3D}~\cite{fomo3d} contract was subjected to such kind of attack.}

\begin{figure}
\centering
\includegraphics[width=8.55cm]{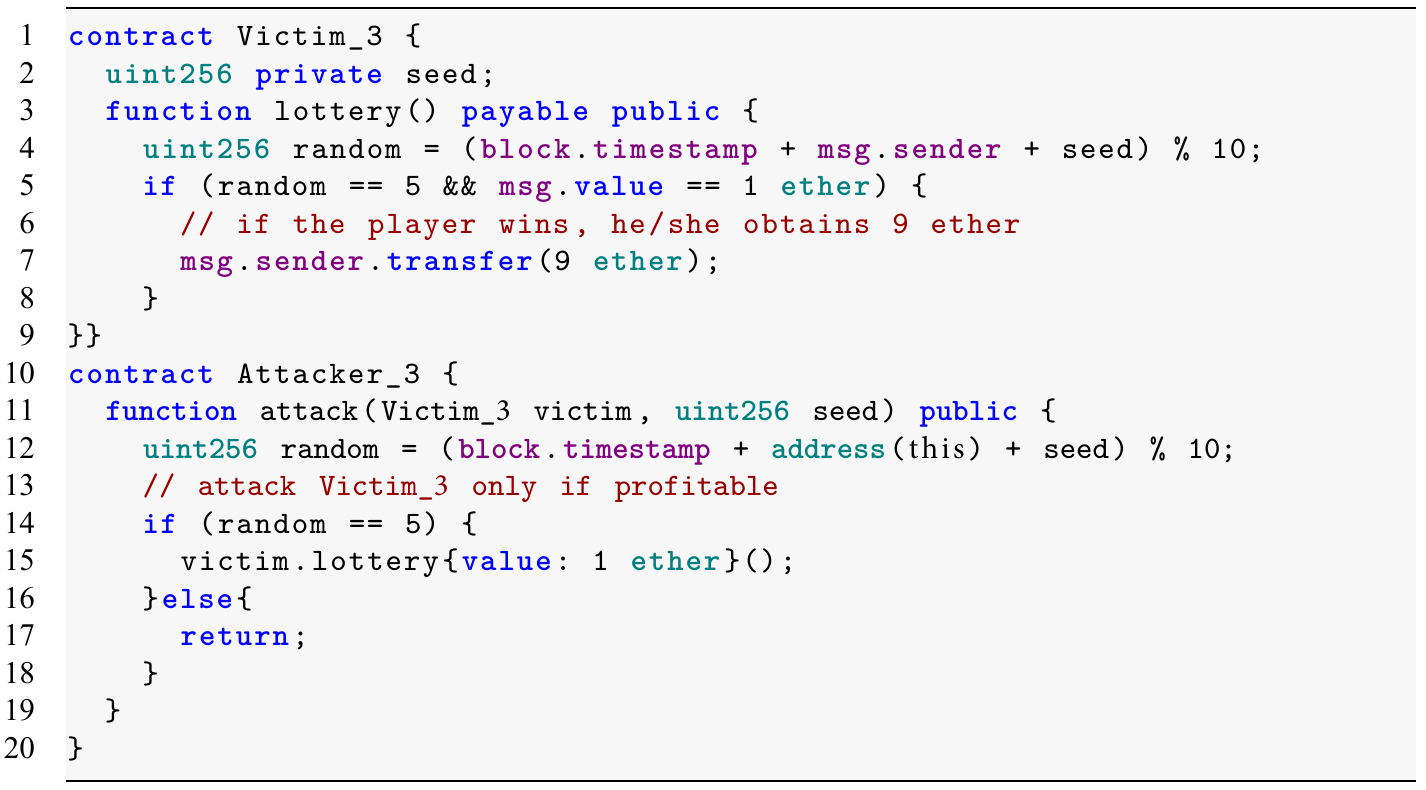}
\vspace{-0.2em}
\caption{{A simplified example to show a direct prediction attack.}} 
\label{fig5}
\vspace{-1.2em}
\end{figure}

\emph{Direct Prediction.}\quad 
Besides manipulating the generation of random number seeds, an attacker can also pre-calculate random numbers before submitting a bet, and then place a bet only when he can make a profit. We present a direct prediction attack example in Fig.~\ref{fig5}. Particularly, if the random number predicted by an attacker exactly matches the lottery number (line 14), the attacker will be rewarded with 9 times the participation fee (line 7). If the attacker fails to make a profit, he/she can stop the transaction early (line 17), thus avoiding losing their capital.

\begin{figure}
\centering
\includegraphics[width=8.55cm]{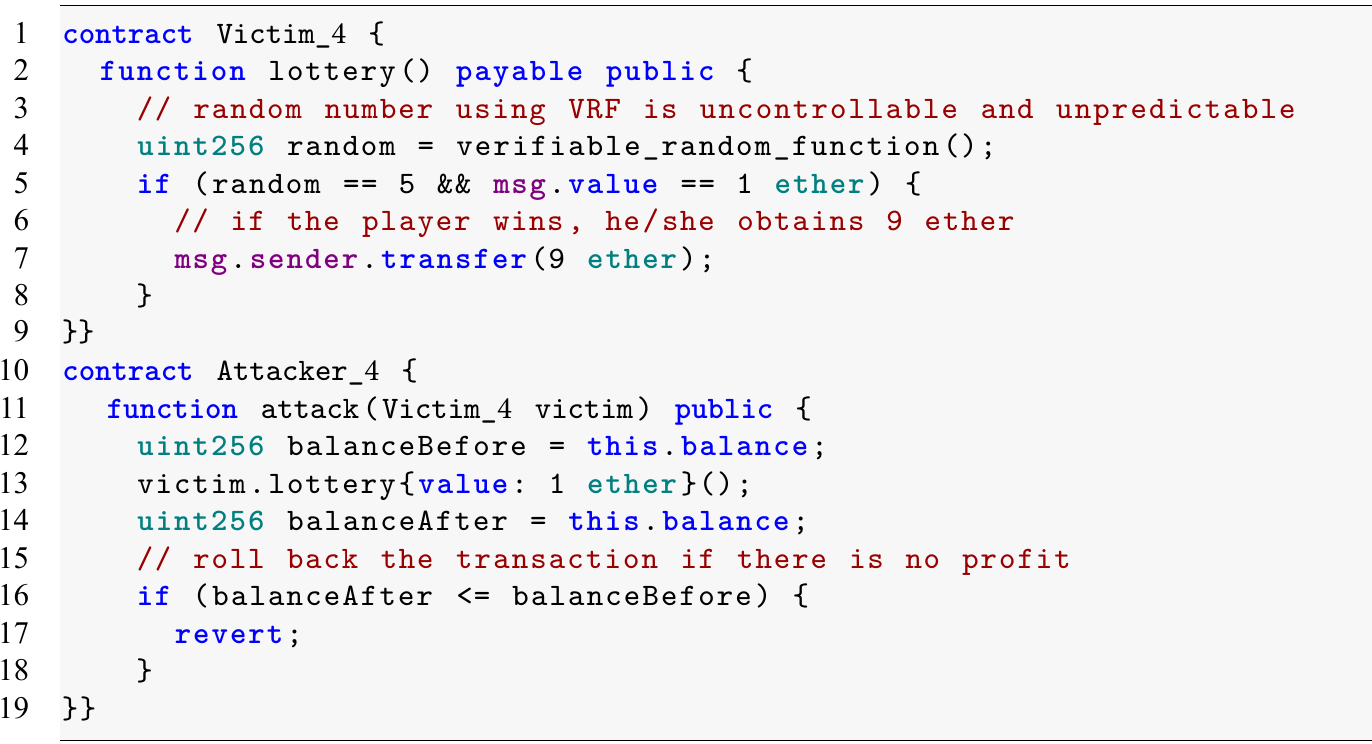}
\vspace{-0.2em}
\caption{A simplified example to show a transaction rollback attack.} 
\label{fig4}
\vspace{-1.2em}
\end{figure}

\emph{Transaction Rollback.}\quad Here, we further show an interesting random number attack. Sometimes the random number generation is complicated and unpredictable, e.g., using VRF to produce a random number. At this time, adversaries can exploit the transaction rollback mechanism to achieve a random number attack. As shown in Fig.~\ref{fig4}, the attacker first checks the balance of his account (line 12). Then, the attacker invokes the function \emph{lottery} of the victim contract \emph{Victim}\_3 (line 13). After that, the attacker checks the account balance again (line 14), and determines whether he/she made a profit (line 16). If not, this transaction will be rolled back to the previous state, and the participation fee is given back to the attacker, thereby avoiding losses. {Note that we have indeed detected many transaction rollback attacks from real-world Ethereum transactions in the experiments.}

\begin{figure*}
\centering
\includegraphics[width=17.2cm]{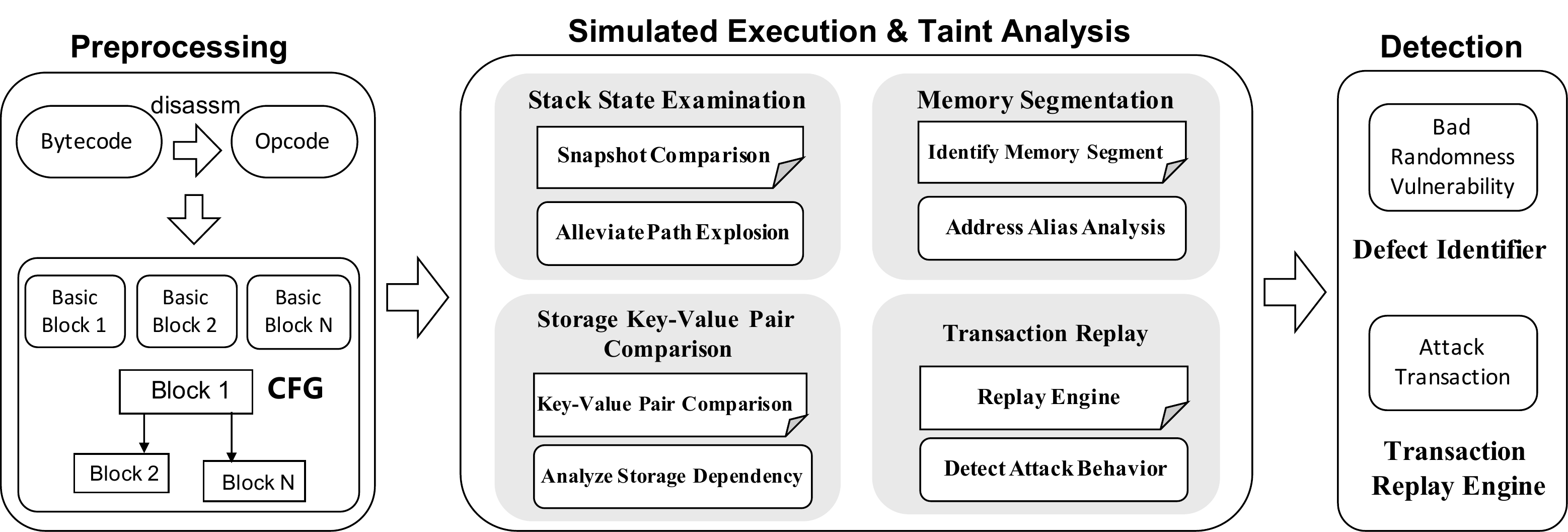}
\vspace{-0.2em}
\caption{A high-level overview of our pipeline. In the preprocessing phase, \emph{RNVulDet} disassembles the bytecode of the smart contract into opcode and splits the opcode into a set of basic blocks for constructing the control flow graph. In the simulated execution phase, \emph{RNVulDet} engages in four main components to facilitate taint analysis, giving the detection result of bad randomness vulnerabilities and random number attack transactions.} 
\label{fig_overview}
\vspace{-1.2em}
\end{figure*}

\subsubsection{Patterns for Detecting Random Number Attack}
\label{patterns_attack}
In this subsection, we further define specific patterns to facilitate random number attack detection. Exactly, we design two types of attack patterns, where one is to identify the random number manipulation or prediction attack, while the other is to discover transaction rollback attack.

(1) {Empirically, we observe that the first three types of random number attacks introduced in \S\ref{classification_attack} can be concluded as \emph{random number manipulation or prediction} attacks because they share a common pattern. That is, }within a transaction, if a \emph{caller} contract adopts the same vulnerable instructions (which will be defined in \S\ref{overview}) of the \emph{callee} contract to generate a pseudo-random number, and the random number generated in the \emph{callee} contract contaminates the parameters of the Ether transfer instruction ({i.e.,} \texttt{CALL}) or the conditional jump instruction ({i.e.,} \texttt{JUMPI}) before the transfer instruction, the \emph{caller} contract is considered to be an \emph{attack} contract and this transaction issued by the \emph{caller} contract is marked as {an attack transaction ({i.e.,} either random number manipulation or prediction).} 

(2) We detect the transaction rollback attack by checking if the rollback transaction and the profit-making transaction occur simultaneously within a short time. Particularly, we identify a rollback transaction by checking whether it has four steps: \ding{172} a \emph{caller} contract issues an invocation to a \emph{callee} contract; \ding{173} the \emph{caller} contract queries its balance; \ding{174} the balance information contaminates the subsequent conditional jump instruction ({i.e.,} \texttt{JUMPI}); \ding{175} the \emph{caller} contract utilizes the \texttt{REVERT} instruction to rollback the current transaction. 

Furthermore, we confirm a profit-making transaction by checking if it contains another four steps: \ding{172} a \emph{caller} contract issues an invocation to a \emph{callee} contract; \ding{173} the \emph{callee} contract transfers money to the \emph{caller} contract; \ding{174} the \emph{caller} contract queries its balance; \ding{175} the balance information contaminates the subsequent conditional jump instruction ({i.e.,} \texttt{JUMPI}).

\section{Our Method}
\label{approach}

\subsection{Overview of RNVulDet}
\label{overview}
\emph{Design Overview.}\quad Before diving into the details of our method, we outline the high-level pipeline of \emph{RNVulDet}, which comprises preprocessing and simulated execution. Fig.~\ref{fig_overview} depicts the overall architecture of our proposed \emph{RNVulDet}. In particular, \emph{RNVulDet} engages in four main components to facilitate taint analysis, {i.e.,} stack state examination, memory segmentation, storage key-value pair comparison, and transaction replay. 

\emph{Preprocessing.}\quad Fundamentally, \emph{RNVulDet} takes the bytecode of smart contracts as input. Solidity source code is also allowed, but it will be compiled to bytecode once it is fed into \emph{RNVulDet}. Bytecode is then disassembled into opcode (short for operation code, also known as instruction code) by utilizing API provided by {Geth}~\cite{geth}. {\emph{RNVulDet} splits the opcode into a set of basic blocks and generates the control flow graph (CFG) of the smart contract, which consists of basic blocks and control flow edges. The basic block usually comprises the sequence of instructions. \emph{RNVulDet} will symbolically execute instructions in each block. One basic block is connected to subsequent basic blocks through a branch instruction in the CFG. In our analysis, we treat the branch instructions (e.g., \texttt{JUMP}, \texttt{JUMPI}) as the sign of the end of a basic block, which indicates that the branch instruction is considered a flag to segment the basic blocks. If the last instruction is a terminal instruction (e.g., \texttt{STOP}, \texttt{REVERT}, and \texttt{RETURN}), the block type is terminal~\cite{chen2021defectchecker}. Note that we actually do not recover the complete control flow graph of the contract bytecode, but rather restore the control flow execution paths from the bytecode, which is easier to implement than recovering a comprehensive control flow graph. For any given contract, \emph{RNVulDet} is capable of extracting all control flow execution paths.}

It is worth mentioning that detecting vulnerabilities on the bytecode level is very important for Ethereum smart contracts. All the bytecode of smart contracts are stored on the blockchain and can be readily accessed, but only a very small fraction (less than $1\%$) of smart contracts are open-sourced~\cite{chen2019tokenscope}. In particular, smart contracts usually invoke external contracts, but the callee contract may not open its source code for inspection. In this context, the caller contract can only identify whether the callee contract is secure through the bytecode.

\emph{Simulated Execution.}\quad 
\emph{RNVulDet} simulates the runtime environment of the Ethereum Virtual Machine (EVM), which symbolically executes each single EVM instruction one at a time. To satisfy the requirements of symbolic execution, \emph{RNVulDet} simulates three kinds of storage structures, \emph{i.e.,} Stack, Memory, and Storage. During the execution, we can obtain the jump relations between basic blocks. \emph{RNVulDet} executes the instructions in the basic blocks sequentially and generates the corresponding instruction instance (which refers to the execution result of instruction) for each instruction code. These instruction instances will be stored in the three types of storage structures. 

\begin{definition}[\textbf{Vulnerable Instruction}]\quad 
Given the collection $\mathbb{T}$ that contains a set of EVM instructions, if any instruction in $\mathbb{T}$ used as a seed to generate pseudo-random numbers may introduce a bad randomness vulnerability, we say that every given instruction $t \in \mathbb{T}$ is vulnerable.
\end{definition}

Specifically, \emph{RNVulDet} executes basic blocks of the CFG in a depth-first traversal fashion. When the simulated execution encounters a \emph{vulnerable instruction}\footnote{Notably, we focus on seven types of vulnerable instructions, namely, \texttt{BLOCKHASH, COINBASE,  DIFFICULTY, GASLIMIT, MOD\_TIME, NUMBER,} and \texttt{TIMESTAMP}. }, \emph{RNVulDet} will mark it as the taint source, which is the headstream of taint propagation. Once the \texttt{CALL} instruction is triggered, the taint sink check will be performed to verify whether the parameters of the \texttt{CALL} instruction and the \texttt{JUMPI} instruction on the current execution path are contaminated by the taint source. Finally, \emph{RNVulDet} analyzes the taint analysis results and reveals a bad randomness vulnerability with the assistance of a pattern analyzer. {Note that \emph{RNVulDet} is able to pinpoint the exact location of each detected bad randomness vulnerability. If a smart contract contains more than one bad randomness vulnerability, \emph{RNVulDet} can reveal all of them as well as their exact locations.} In the following, we show an example of how \emph{RNVulDet} uses specific patterns to expose the bad randomness vulnerability.

\emph{Vulnerability Pattern.}\quad 
\emph{RNVulDet} considers that a smart contract possesses a bad randomness vulnerability if the random number generated by \emph{vulnerable instructions} contaminates the Ether transfer operation ({i.e.,} \texttt{CALL} instruction). Specifically, we design three patterns to expose a bad randomness vulnerability. 
The first pattern \texttt{CALLJUMPI} checks if the \emph{vulnerable instruction} affects the triggering condition of the transfer operation, {namely} whether the \emph{vulnerable instruction} contaminates the \texttt{JUMPI} instruction which appears on the execution path of the \texttt{CALL} instruction.
The second pattern is \texttt{CALLToAddress} which checks if the \emph{vulnerable instruction} contaminates the transfer address, {i.e.,} the parameter \emph{ToAddress} of \texttt{CALL} instruction. 
The third pattern \texttt{CALLValue} concerns whether the \emph{vulnerable instruction} contaminates the transfer amount, {i.e.,} the parameter \emph{Value} of \texttt{CALL} instruction. 
\emph{RNVulDet} reports that a function has a bad randomness vulnerability if it fulfills the combined pattern: $\texttt{CALLJUMPI} \wedge \texttt{CALLToAddress} \wedge \texttt{CALLValue}$.

\subsection{Stack State Examination}
\label{sec:stack}
Existing symbolic execution methods tend to traverse all execution paths as much as possible to expose possible vulnerabilities in smart contracts. However, an exhaustive traversal strategy introduces the inherent problem of path explosion, resulting in high performance overhead. Therefore, we consider proposing a method to alleviate the path explosion problem, which aims to reduce the execution of potential repeated program paths during taint analysis, improving the overall efficiency of simulated execution.

\vspace{2px}
\textbf{Rule 1: Snapshot Comparison.}\quad \emph{Given a snapshot $\mathcal{S}_1$ and a historical snapshot $\mathcal{S}_h$, we say that $\mathcal{S}_1$ is equal to $\mathcal{S}_h$ \emph{iff} all jump addresses and taint sources in $\mathcal{S}_1$ are identical to those of $\mathcal{S}_h$.}
\vspace{2px}

Technically, we design a stack state examination mechanism to prune potential invalid execution paths. \emph{RNVulDet} will store the stack state snapshot for each basic block. Whenever the simulated execution reaches a basic block, \emph{RNVulDet} performs the snapshot comparison (\textbf{Rule 1}) to determine whether the stack state snapshot of the block is equal to the one in the historical stack state snapshots. If it is matched, we consider that continuing along the current program path will not expose new vulnerabilities, thus the execution of this basic block and its subsequent paths will be skipped. Otherwise, the execution will continue along the current program trace.
As such, \emph{RNVulDet} can filter out possible repeated and bug-free paths, enforcing the simulated executor to trigger new traces and unearth potential bugs.

\subsection{Memory Segmentation}
\label{sec:memory}
In practice, \emph{RNVulDet} simulates the executing environment of Memory to analyze memory data dependencies, \emph{namely} read and write memory operands. However, since the operands of {read} and {write} instructions in Memory both are abstract instruction instances (\emph{a.k.a.} the execution results of instructions), it is very difficult to judge whether the memory address accessed by {read} instruction is identical to that of write instruction. This will cause the {read} instruction to be unable to determine which instruction instance of a {write} instruction in Memory is its operand. It is worth noting that this is exactly the classical alias analysis puzzle in the field of program analysis. To address this problem, we propose a memory segmentation paradigm for handling {read} and {write} instruction dependencies in Memory. 

To better understand the principle of memory segmentation, we first introduce how Solidity manages memory. Memory in the EVM is essentially an infinite array, which is addressed by bytes. In the traditional computer, the operating system is responsible for maintaining unused memory, allocating memory, and keeping track of memory that has been used. Unfortunately, there is no operating system in EVM, and thus memory allocation has to be completed by the Solidity compiler. Solidity currently adopts the free memory pointer (FMP) to mark the start of available addresses in Memory. Solidity always places new objects at the free memory pointer and memory is never freed, that is, the smart contract written by Solidity will not reclaim the memory that has already been allocated. As such, Solidity only allocates memory in segments incrementally. Once a new allocation demand occurs,  it will allocate a section of memory space after the current FMP. Note that the initial FMP is stored in the memory space of \texttt{0x40}.

Inspired by this, we design a memory segmentation-based data dependency analysis paradigm. The core idea is to divide memory into several segments and map each {read} and {write} instruction to the corresponding memory segment one by one. When executing a {read} instruction, we take all instances of {write} instructions in the associated memory segment as the operands of the {read} instruction. Thereafter, the problem that we need to solve is how to mark each memory segment and how to identify the memory segment corresponding to {read} and {write} instructions.

\emph{Memory Segment Identification.}\quad  According to the memory management strategy of Solidity, the FMP will be updated every time memory is allocated. We set the segment identifier at the location where each FMP is updated, respectively. In the meantime, whenever a memory allocation occurs, the write memory instruction \texttt{MSTORE} is executed. Therefore, we also set a unique identifier for each \texttt{MSTORE}. {Moreover, to satisfy the memory allocation requirement, Solidity will allocate a section of memory after the current FMP.} We can observe that the operand address of each {read} and {write} instruction instance is represented in the form of FMP + $\Delta$, where $\Delta$ represents the offset of the address of the written operand from that of the current FMP. {Notably, the value of $\Delta$ is determined according to the specific location of the address to be written.} As a result, we are able to utilize the FMP identifier and relevant offset to determine which memory segment the {read} and {write} instructions correspond to.

\begin{figure}
\centering
\includegraphics[width=8.6cm]{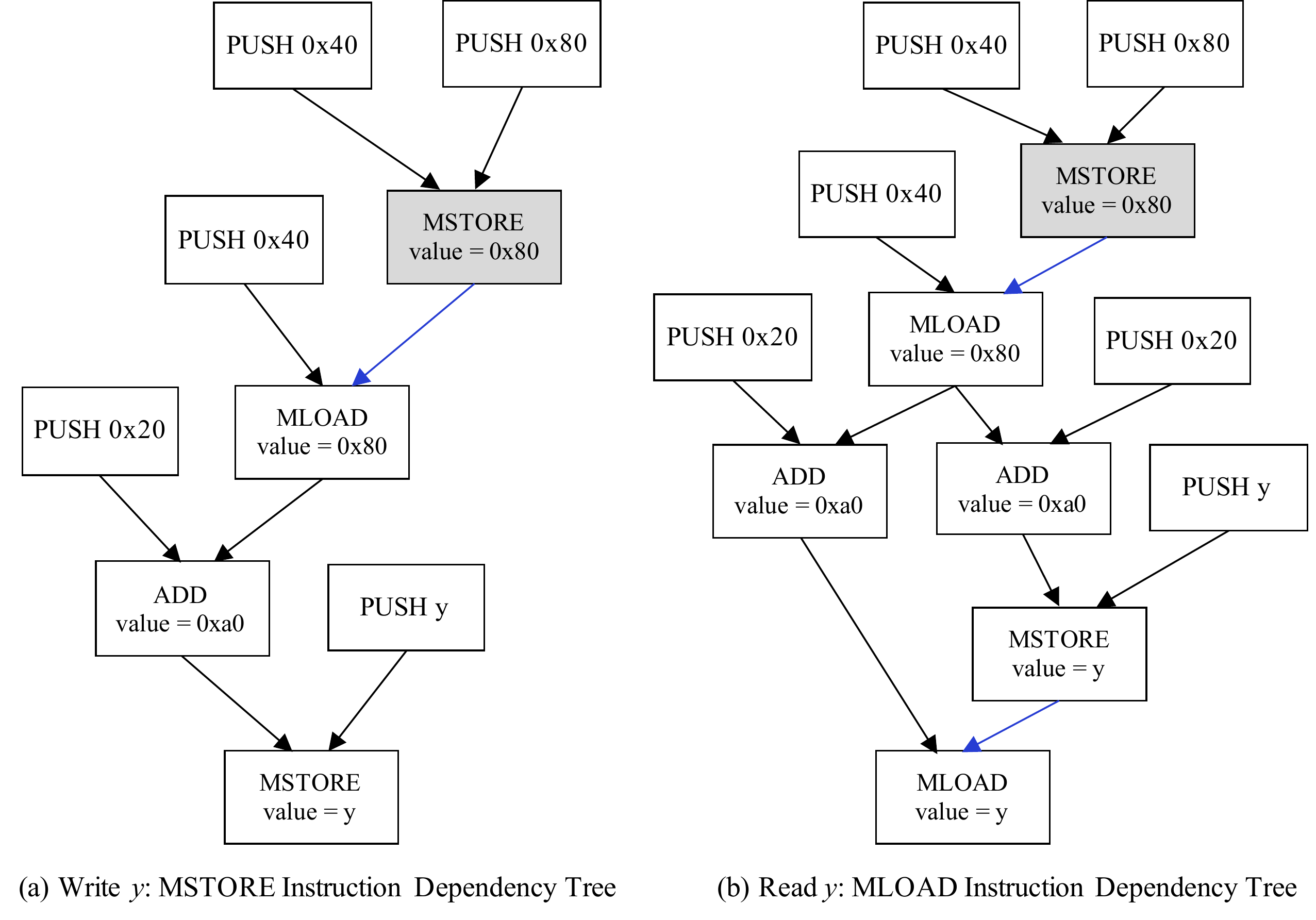}
\vspace{-0.2em}
\caption{ Example: Data dependency analysis of MSTORE and MLOAD instructions in Memory. Node denotes the instruction instance. An arrow from node A to B indicates that A is the operand of B, where ``$\rightarrow$'' represents the stack operand and ``${\color{blue} \rightarrow}$'' denotes the memory operand.} 
\label{fig_memory}
\vspace{-1.2em}
\end{figure}

\textit{Memory Read and Write Dependency Analysis.}\quad Fig.~\ref{fig_memory} presents a specific example of memory {read} \& {write} dependency analysis in \emph{RNVulDet}. Technically, \emph{RNVulDet} constructs the instruction dependency trees according to the address operand of read and write instruction instances in Memory, thereby determining which memory segment the instruction instance is associated with. For \texttt{MSTORE} instruction dependency tree, \emph{RNVulDet} first seeks out the \texttt{MSTORE} instruction with \texttt{0x40} as the address parameter, and then identifies the unique identifier of \texttt{MSTORE} instruction instance, so as to find the corresponding memory segment in Memory. Thereafter, \emph{RNVulDet} places the instance of \texttt{MSTORE} instruction that writes value $y$ in the memory segment. For \texttt{MLOAD} instruction dependency tree, \emph{RNVulDet} finds the corresponding memory segment in Memory according to the unique identifier of \texttt{MSTORE} instruction instance, and then fetches all instruction instances in the memory segment as the memory operands of \texttt{MLOAD} instruction. Finally, \emph{RNVulDet} extracts taint sources of the memory operand of \texttt{MLOAD}  instruction instance that reads value $y$, thereby realizing the taint propagation in Memory.

\subsection{Storage Key-Value Pair Comparison}
\label{sec:storage}
Storage is organized as a key-value store, in which the instructions for reading and writing Storage are \texttt{SLOAD} and \texttt{SSTORE}, respectively. For analyzing data dependency in Storage, it is critical to determine whether the keys of \texttt{SLOAD} and \texttt{SSTORE} are identical. {Yet, this is by no means easy.} In addition, since data in Storage will be permanently stored in the Ethereum world state, it also involves the problem of cross-transaction taint propagation.

\begin{figure}
\centering
\includegraphics[width=8.665cm]{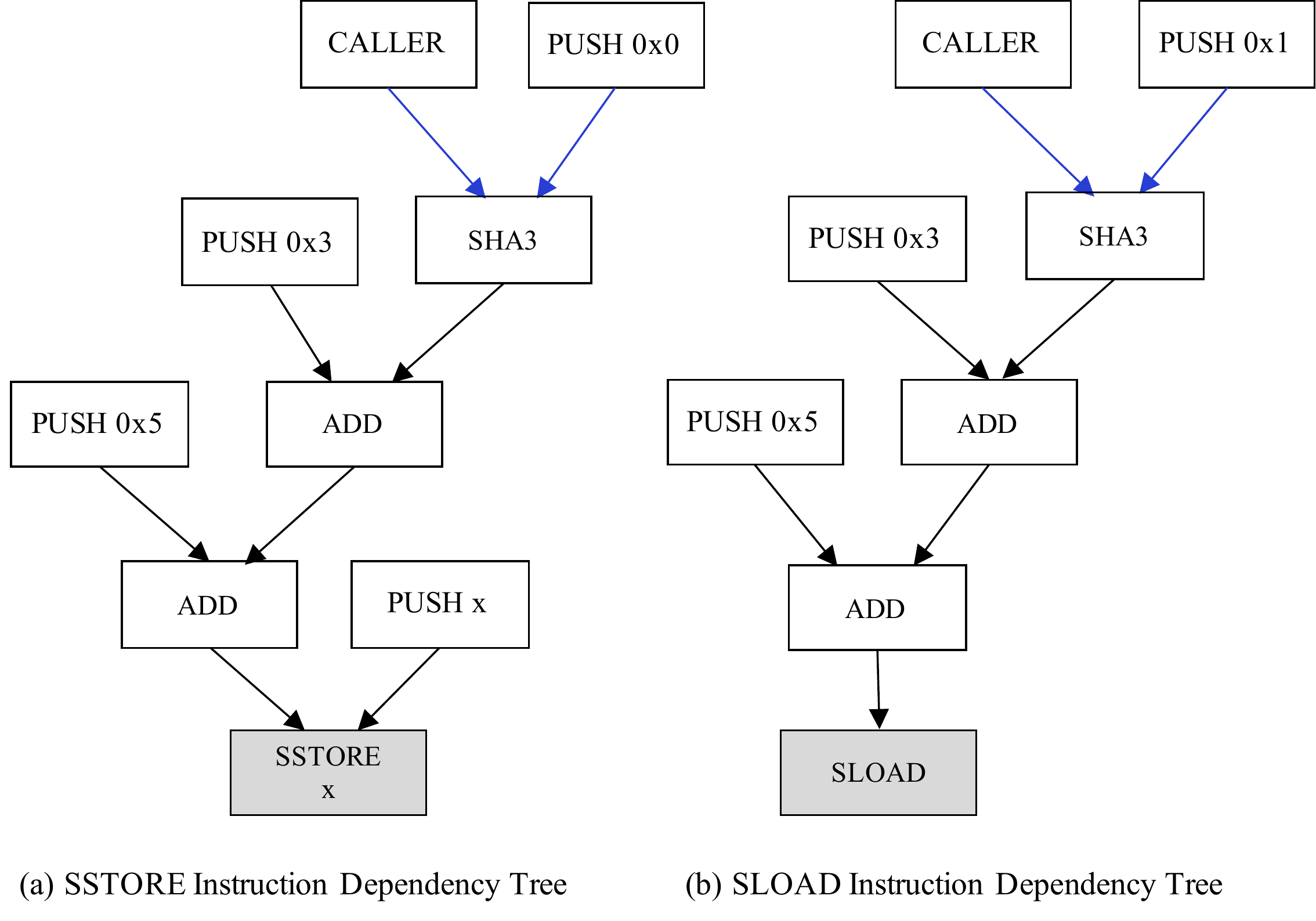}
\vspace{-0.2em}
\caption{ Example: Data dependency analysis of SSTORE and SLOAD in Storage.} 
\label{fig6}
\vspace{-1.2em}
\end{figure}

\emph{Key-Value Pair Comparison.}\quad We propose a novel key-value pair comparison strategy to address the data dependency analysis in Storage. During the simulation execution, whenever encountering the \texttt{SSTORE} instruction, \emph{RNVulDet} will add the instruction instance to the end of Storage. When \texttt{SLOAD} is executed, \emph{RNVulDet} searches forward from the end of Storage to find whether there is an instruction instance of \texttt{SSTORE} and whether the key of \texttt{SSTORE} is equal to that of \texttt{SLOAD}. Assuming there exists an instruction instance of \texttt{SSTORE} that has the same key as the instruction \texttt{SLOAD}, the instruction instance of \texttt{SSTORE} is used as the operand of \texttt{SLOAD}. Particularly, if multiple different \texttt{SSTORE} instructions execute on the same key, the value of the later \texttt{SSTORE} instruction will overwrite the value of the previous \texttt{SSTORE} instruction. Therefore, for the \texttt{SLOAD} instruction, it only needs to read the latest value of the key. 

In practice, we found that the key of Storage represents a polynomial expression in the form of the sum of multiple instruction instances. Therefore, to facilitate the data dependency analysis of \texttt{SSTORE} and \texttt{SLOAD} in Storage, we consider flattening the instruction dependencies of the keys of \texttt{SSTORE} and \texttt{SLOAD} instructions into a \emph{polynomial form}. In this context, determining whether the keys of \texttt{SSTORE} and \texttt{SLOAD} are equal in Storage is to judge whether each item of the \texttt{SSTORE} and \texttt{SLOAD} polynomials is a one-to-one correspondence.
Note that the polynomial also contains function calls. The function with the same arguments may return different results since the values of some instructions can be changed during the simulation execution, e.g., \texttt{PC}, \texttt{MSIZE}, and \texttt{GAS}. Encouragingly, all the function calls in the polynomial will be included during the simulated execution. Therefore, in most cases, \emph{RNVulDet} can ensure that when the polynomials of \texttt{SSTORE} and \texttt{SLOAD} are the same, they can return the same result.

Fig.~\ref{fig6} presents a specific example to compare the key between the \texttt{SSTORE} instruction and the \texttt{SLOAD} instruction. \emph{RNVulDet} constructs the instruction dependency trees of \texttt{SSTORE} and \texttt{SLOAD}, respectively. From the figure, we can observe that although the addition result (\emph{i.e.,} 0x3+0x5=0x8) of the immediate value in the \texttt{SSTORE} instruction dependency tree is the same as that of the \texttt{SLOAD}, the operands of the hash instruction \texttt{SHA3} are \texttt{CALLER} and \texttt{PUSH 0x0} in \texttt{SSTORE} instruction tree while \texttt{CALLER} and \texttt{PUSH 0x1} in \texttt{SLOAD} instruction tree. Therefore, \emph{RNVulDet} can confirm that the key of the \texttt{SLOAD} instruction is different from that of the \texttt{SSTORE} instruction.

\emph{Cross-Transaction Taint Propagation.}\quad 
Unlike Stack and Memory which will reset the storage space after one contract execution, Storage will store data in the Ethereum world state for a long time. Therefore, when analyzing the data dependency of Storage, we do not only deal with taint analysis in the current contract execution but also further consider the problem of cross-transaction taint propagation in possible subsequent executions. For example, in the real-world game contract \emph{FoMo3D}~\cite{fomo3d}, when a user calls \emph{FoMo3D} to play the game at the first invocation, the contract will record whether the user wins based on the matching result between the user’s guess and the generated pseudo-random number of \emph{FoMo3D}. If the user guesses it and calls \emph{FoMo3D} to withdraw the reward at the second invocation, the contract determines whether to transfer the reward to the user according to matching results stored in Storage.

In this context, when the user calls the \emph{FoMo3D} game contract firstly, the random number does not yet directly contaminate the \emph{transfer} operation, but affects the matching result which will be stored in Storage. When the user invokes \emph{FoMo3D} secondly, the matching result stored in Storage interferes with the \emph{transfer} operation. Therefore, if the \emph{FoMo3D} contract is only executed once by \emph{RNVulDet}, the bad randomness vulnerability will not be exposed. To deal with such a problem, \emph{RNVulDet} considers performing the contract many times. After every execution, the storage key-value pairs contaminated by \emph{vulnerable instructions} will be retained, so that the subsequent execution could realize the cross-transaction taint propagation in Storage.

\subsection{Transaction Replay}
\label{sec:transaction}
In this stage, \emph{RNVulDet} turns to random number attack monitoring, which engages in a transaction replay technique to detect potential attack transactions. 

To monitor the transactions at the runtime, \emph{RNVulDet} engages in a replay engine that is capable of replaying arbitrary transactions on Ethereum. Following our previous work~\cite{wu2022time}, we define corresponding rules to locate suspicious random number attack transactions, and more importantly, provide fine-grained states that are needed by the replay engine. 
After retrieving the historical states, the replay engine re-executes these suspicious transactions. During the execution, \emph{RNVulDet} adopts the dynamic taint analysis technique to introspect the execution of transactions and monitor the runtime states of a smart contract, thereby determining whether the transaction possesses a random number attack behavior.

\begin{algorithm}[t]
   \footnotesize    
    \setstretch{0.9} 
  \caption{ \textsc{Dynamic Taint Analysis} }
  \label{alg1}

  \KwIn{ $C_{target}$, $T_{suspicious}$, $C_{caller}$, $S_{vulnerable}$ }
  \KwOut{ $T_{attack}$ }

$executor.setup(T_{suspicious}.readSet, T_{suspicious}.input);$

${taint}_{caller} \leftarrow set(\,);$

${taint}_{target} \leftarrow set(\,);$

\While{$executor.hasNextInst(\,)$}{
  $inst \leftarrow executor.getNextInst(\,);$

  \If{$inst \in S_{vulnerable}$}{$taintSource(inst);$}

  $taintPropagate(inst);$

  \If{$inst == \textrm{JUMPI} \vee inst == \textrm{CALL}$}{$taintSink(inst);$}

}

\If{$taint_{caller} \cap taint_{target} \neq \varnothing $}{$T_{attack} \leftarrow T_{suspicious};$ }

\BlankLine

  \SetKwFunction{FtaintSink}{$taintSink$}

  \SetKwProg{Fn}{Function}{:}{}
  \Fn{\FtaintSink{$inst$}}{

        \If{$inst == \textrm{JUMPI}$}{
          \If{$executor.currentContract == C_{caller}$} {
            $taint_{caller}.add(inst.taints);$
          }\Else {
            $taint_{target}.add(inst.taints);$
          }
        }

        \If{$inst == \textrm{CALL}$} {
          \If{$inst.from == C_{caller} \wedge inst.to == C_{target}$}{
              $taint_{caller}.add(inst.taints);$
          }
          \If{$inst.from == C_{target} \wedge inst.to == C_{caller}$}{
              $taint_{target}.add(inst.taints)$
          }
        }
  }

\end{algorithm}

{Following the attack patterns defined in \S\ref{patterns_attack}, we introduce a dynamic taint analysis method to detect corresponding random number attacks, which is summarized in Algorithm~\ref{alg1}.}
{\emph{RNVulDet} takes the target contract $C_{target}$, the suspicious transaction $T_{suspicious}$, the caller contract $C_{caller}$, and the collection $S_{vulnerable}$ that contains a set of \emph{vulnerable instructions} as inputs}, and then outputs the marked attack transaction $T_{attack}$. \emph{RNVulDet} will initialize the taint aggregation set of the caller contract and the target contract as ${taint}_{caller}$ and ${taint}_{target}$, respectively.
First, \emph{RNVulDet} reads a collection of suspicious transactions from the Ethereum world state database and restores the execution environment of the transactions (line 1). Then, the loop from lines 4 to 10 marks taint sources and performs taint propagation. \emph{RNVulDet} executes the contract instructions sequentially and continuously obtains new instructions through \emph{executor.getNextInst()} (line 5). {If an instruction belongs to $S_{vulnerable}$, the instruction is vulnerable and will be marked as a taint source, \emph{i.e., taintSource(inst)} (lines 6-7)}. The taint source will propagate as the sequence of program instructions is executed. When encountering a \texttt{JUMPI} instruction or a \texttt{CALL} instruction, the taint sink check is triggered, \emph{i.e., taintSink(inst)} (lines 9-10). When the execution of the suspicious transaction is completed, \emph{RNVulDet} computes the intersection between ${taint}_{caller}$ and ${taint}_{target}$. If the intersection is not empty, \emph{RNVulDet} will mark the suspicious transaction $T_{suspicious}$ as the real attack transaction $T_{attack}$.

More specifically, for instructions \texttt{JUMPI} and \texttt{CALL}, \emph{RNVulDet} adopts different taint sink check strategies, respectively. 
{(1)} If the \texttt{JUMPI} instruction exists in the caller contract $C_{caller}$, the taint source \emph{inst.taints} that contaminates this instruction is added to ${taint}_{caller}$ (lines 15-16), which indicates that the control flow of the caller contract is affected by \emph{vulnerable instructions}. Conversely, if the instruction exists in the target contract $C_{target}$, \emph{inst.taints} will be added to ${taint}_{target}$ (line 18), which means that the control flow of the target contract is influenced by $S_{vulnerable}$.
{(2)} If the \texttt{CALL} instruction indicates that $C_{caller}$ invokes $C_{target}$, the taint source \emph{inst.taints} is added to ${taint}_{caller}$ (lines 20-21), which implies that $C_{caller}$ uses the pseudo-random number generated by \emph{vulnerable instructions} to call ${taint}_{target}$. On the contrary, if the \texttt{CALL} instruction represents that ${taint}_{target}$ calls $C_{caller}$, \emph{inst.taints} will be added to ${taint}_{target}$ (lines 22-23), which suggests that the transfer of ${taint}_{target}$ to $C_{caller}$ is influenced by \emph{vulnerable instructions}. 

To be more specific, the dynamic taint analysis technique is fully applicable to the detection of the \emph{random number manipulation or prediction} attack ({i.e.,} the first three types of random number attacks described in \S\ref{classification_attack}). For the \emph{transaction rollback} attack, \emph{RNVulDet} will confirm if a transaction contains the four steps based on the pattern defined in \S\ref{patterns_attack}, where the fourth step, \emph{namely,} whether the balance information contaminates the conditional jump instruction (i.e., \texttt{JUMPI}), is determined by using Algorithm~\ref{alg1}.

\section{Experiment}
\label{evaluation}
In this section, we carry out extensive experiments to evaluate the performance of \emph{RNVulDet}, seeking to address the following research questions.
\begin{itemize}[topsep=2pt, leftmargin=\dimexpr\labelwidth + 2 \labelsep\relax]
\item \textbf{RQ1:} Can \emph{RNVulDet} effectively identify {bad randomness} vulnerabilities in Ethereum smart contracts? How is its performance against state-of-the-art tools?
\item \textbf{RQ2:} How is the execution efficiency of \emph{RNVulDet} in analyzing smart contracts compared with other methods?
\item \textbf{RQ3:} How does \emph{RNVulDet} perform in detecting random number attack transactions?
\end{itemize}
In what follows, we first introduce the experimental settings, followed by answering these questions one by one.

\subsection{Experimental Settings}
\label{sec:experimental_setup}

\emph{Datasets.}\quad For this study, we crafted three kinds of datasets for validating the usefulness and effectiveness of \emph{RNVulDet} from multiple perspectives.

\emph{(1) Dataset\_1.}\quad We first collected Ethereum smart contracts that have bad randomness vulnerabilities from different sources, including Ethereum official platform, CVE database, GitHub repositories, and blog posts that analyze smart contracts. In total, we collected 34 smart contracts and manually confirmed that they indeed possess bad randomness vulnerabilities. {Note that a smart contract may contain more than one bad randomness vulnerability.} We use this dataset to evaluate the false negatives of the detection tools we studied.

\emph{(2) Dataset\_2.}\quad We rounded up popular Ethereum smart contracts from \texttt{DappRadar}~\cite{dappRadar} in February 2022. Specifically, we screened out the top 100 decentralized applications according to the user transaction volume at that time. {As a result, we collected a total of 214 popular smart contracts. Note that these popular contracts have been rigorously audited before launch. We thus default the 214 smart contracts that are free of bad randomness vulnerabilities. Then, in the experiments, we manually check the contracts that are reported to have a bad randomness vulnerability by each tested tool, so as to assess the false positives of the corresponding detection tools.}

\emph{(3) Dataset\_3.}\quad We collected historical transactions of smart contracts in the first 12 million blocks in Ethereum, with block timestamps ranging from July 2015 to May 2021. Explicitly, this dataset contains rich information of each transaction, including transaction hash, transaction sender address, transaction value, and calling parameters. We also recorded all world states during each transaction execution, such as the account balance of each block. In total, we collected 38,665,879 contract bytecode, as well as the corresponding addresses and historical transactions. With the help of our previous tool~\cite{wu2022time}, we filtered out 4,617 potential victim contracts, 43,051 potential malicious contracts, and 49,951 suspicious transactions for experiments.

\emph{Baseline Tools.}\quad After empirically scrutinizing existing vulnerability detection tools for smart contracts, we found that two industrial-grade static tools could detect weak randomness vulnerabilities, i.e., \emph{Slither} and \emph{Mythril}. We select \emph{Slither} v0.8.2 and \emph{Mythril} v0.22.35. {Slither}~\cite{feist2019slither} is a static analysis framework that converts smart contracts into an intermediate representation and applies static program analysis techniques to discover vulnerabilities. {Mythril}~\cite{mythril} is a static tool that relies on concolic analysis, taint analysis, and control flow analysis to detect smart contract vulnerabilities. Moreover, we also compare with one dynamic tool, i.e., \emph{ConFuzzius}, which can detect the block dependency vulnerability, a main category of bad randomness vulnerabilities. ConFuzzius~\cite{torres2021confuzzius} adopts a hybrid fuzzing strategy for smart contract vulnerability detection. 
While the baseline tools are able to take care of multiple types of vulnerabilities, they merely identify a subset of bad randomness vulnerabilities, lacking detailed and fine-grained analysis of random number vulnerabilities and attacks. Furthermore, while the baseline tools can detect bad randomness vulnerabilities, they still struggle to detect random number attacks.

\emph{Evaluation Metrics.}\quad There are five measurements obtained from our experiments to demonstrate the performance of tools, i.e., True Positive (TP), True Negative (TN), False Positive (FP), False Negative (FN), and CPU running time. TP represents the number of data items for which the detection tool correctly predicts samples with vulnerabilities. TN is the number of samples that the tool successfully detected without vulnerabilities. FP represents the number of samples that were falsely detected as having vulnerabilities. FN refers to the number of samples that were wrongly detected as projects without vulnerabilities. CPU running time indicates the CPU time spent to analyze the contracts.

All experiments were performed on a computer running Ubuntu 21.04 Server and equipped with two Intel(R) Xeon(R) Silver 4214 CPUs at 2.20GHz and 128GB of memory. For a fair comparison, we set a timeout of 3,600 seconds for each tool. When it comes to timeout, the vulnerability detection will be stopped while the vulnerabilities already found before the timeout will not be lost.

\subsection{Experimental Results and Analysis}
\label{sec:results}

\subsubsection{Effectiveness (RQ1)}
First, we benchmark \emph{RNVulDet} against existing vulnerability detectors on \emph{Dataset\_1}, which consists of 34 smart contracts that have bad randomness vulnerabilities. We present the true positives and false negatives of each tool. Experimental results are summarized in Table~\ref{table2}. {From the table, we can observe that \emph{RNVulDet} significantly outperforms other tools, and existing methods have not yet achieved a high true positive rate. For example, for static tools, \emph{RNVulDet} generates 30 true positives, 12 and 14 more than \emph{Slither} and \emph{Mythril} respectively, while yielding only 4 false negatives. 
Furthermore, the dynamic tool \emph{ConFuzzius} only generates 7 true positives, 23 less than \emph{RNVulDet}, respectively. Intuitively, for fuzzing tools, if they do not cover some program regions, they will not reveal bugs in those regions. It is worth noting that \emph{ConFuzzius} experienced an error on 15 contracts of \emph{Dataset\_1} due to a problem within its contract deployment mechanism.}

\renewcommand{\arraystretch}{1.05}
\begin{table}
	\centering
	\caption{ {True positive and false negative of each tool on Dataset\_1, where false negative represents that a tool fails to reveal a bad randomness vulnerability.} }
	\vspace{-0.9em}
	\resizebox{0.48\textwidth}{!}{
		\begin{tabular}{cccc}
			\bottomrule 
			\textbf{Method} & \textbf{True Positive (TP)} & \textbf{False Negative (FN)} & \textbf{Error} \\
			\hline
			 Slither~\cite{feist2019slither}  & 18  & 16 & 0 \\
			 Mythril~\cite{mythril} & 16  & 18  & 0 \\
			 {ConFuzzius~\cite{torres2021confuzzius}} & {7}  & {12}  & {15} \\
			 \textbf{RNVulDet} & \textbf{30} & \textbf{4} & \textbf{0} \\
			 \toprule
	\end{tabular} 
	} 
\label{table2}
\vspace{-0.8em}
\end{table}

\renewcommand{\arraystretch}{1.05}
\begin{table}
	\centering
	\caption{ {False positive and true negative of each tool on Dataset\_2, where false positive represents that a tool falsely reports a bad randomness vulnerability.} }
	\vspace{-0.9em}
	\resizebox{0.48\textwidth}{!}{
		\begin{tabular}{cccc}
			\bottomrule 
			\textbf{Method} & \textbf{False Positive (FP)} & \textbf{True Negative (TN)} & \textbf{Error} \\
			\hline
			 Slither~\cite{feist2019slither}  & 21  & 187 & 6 \\
			 Mythril~\cite{mythril} & 29  & 185  & 0 \\
			 {ConFuzzius~\cite{torres2021confuzzius}} & {9}  & {53}  & {152} \\
			 \textbf{RNVulDet} & \textbf{4} & \textbf{210} & \textbf{0} \\
		 	 \toprule
	\end{tabular} 
	} 
\label{table3}
\vspace{-1.2em}
\end{table}

Further, we analyze the reasons for the 4 false negatives caused by \emph{RNVulDet}. Among them, 2 false negatives are because the contract does not use \emph{vulnerable instructions} to generate pseudo-random numbers while instead using simple arithmetic instructions, such as \texttt{ADD}, \texttt{SUB}, \texttt{MUL}, and \texttt{DIV}. Since these instructions are normal arithmetic instructions, our tool does not treat them as taint sources, thus reporting false negatives. The other 2 false negatives are because the generated pseudo-random number does not affect the Ether transfer. Explicitly, one of the false negatives is due to the fact that the generated pseudo-random numbers are used to mint NFTs (\emph{a.k.a} Non-fungible tokens)~\cite{nft}, and the other is that the generated pseudo-random number affects other types of tokens rather than Ether. In summary, our system generates the 4 false negatives due to the fact that they fail to match the bad randomness vulnerability detection pattern, i.e., \emph{``vulnerability instructions contaminate the Ether transfer"} (see \emph{Vulnerability Pattern} in \S\ref{overview}).

To further evaluate the effectiveness of \emph{RNVulDet}, we examine it on \emph{Dataset\_2}, which consists of 214 smart contracts that are considered free of bad randomness vulnerabilities. Table~\ref{table3} demonstrates the number of false positives and true negatives produced by each tool, as well as the cases where the tool triggers a runtime error. Specifically, \emph{RNVulDet} only generates 4 false positives, 21 and 29 less than \emph{Slither} and \emph{Mythril}. Moreover, \emph{Slither} ran into a runtime error on 6 smart contracts. Through the manual review, we found that these errors are attributed to the internal breakdown of \emph{Slither} rather than the problem of the experimental settings. {For the dynamic tool, \emph{ConFuzzius} produces 9 false positives. Since fuzzers detect vulnerabilities only when triggering abnormal results during fuzzing, \emph{ConFuzzius} thus reports relatively few false positives. \emph{ConFuzzius} also ran into errors on 153 contracts due to the contract deployment problem.} 

We also investigate the reasons behind the 4 false positives of our tool. Particularly, 2 false positives are because the contract uses a \emph{future block hash} to generate pseudo-random numbers. It is worth mentioning that using a future block hash for pseudo-random number generation is generally considered safe. Since no one can know the content of a future block beforehand, it is almost impossible for miners to predict the pseudo-random number generated by a 256-bit future block hash. Our system produces false positives because \texttt{BLOCKHASH} is a vulnerable instruction. The other 2 false positives are because the contract uses a hybrid random number generation using \emph{``vulnerable instructions + Commit-Reveal"}, which is immune to tampering by attackers or miners. Since the vulnerable instructions are included in the hybrid scheme, our system generates false positives. {It is worth noting that some contracts without random number generators are mostly identified as true negatives. In addition, a pseudo-random number generated by using a verifiable delay function or a hybrid method such as ``commit-reveal + verifiable random function" will also be considered as a true negative.}

\renewcommand{\arraystretch}{1.4}
\begin{table}
        \small
	\centering
	\caption{ CPU running time of each method in analyzing smart contracts. }
	\vspace{-0.9em}
	\resizebox{0.485\textwidth}{!}{
		\begin{tabular}{ccccc}
			\bottomrule 
			\textbf{Method} & \textbf{Average} & \textbf{Lower Quartile} & \textbf{Median} & \textbf{Upper Quartile} \\
			\hline
			 Slither~\cite{feist2019slither}  & 3.34  & 0.79 & 1.41 & 2.11 \\
			 Mythril~\cite{mythril} & 2,009.02  & 81.57  & 2,785.44 & 3,595.48 \\
			 \textbf{RNVulDet} & \textbf{2.98} & \textbf{0.06} & \textbf{0.07} & \textbf{0.12} \\
			 \toprule 
	\end{tabular} 
	} 
\label{table4}
\vspace{-1.2em}
\end{table}

\begin{figure}
\centering
\includegraphics[width=8.4cm]{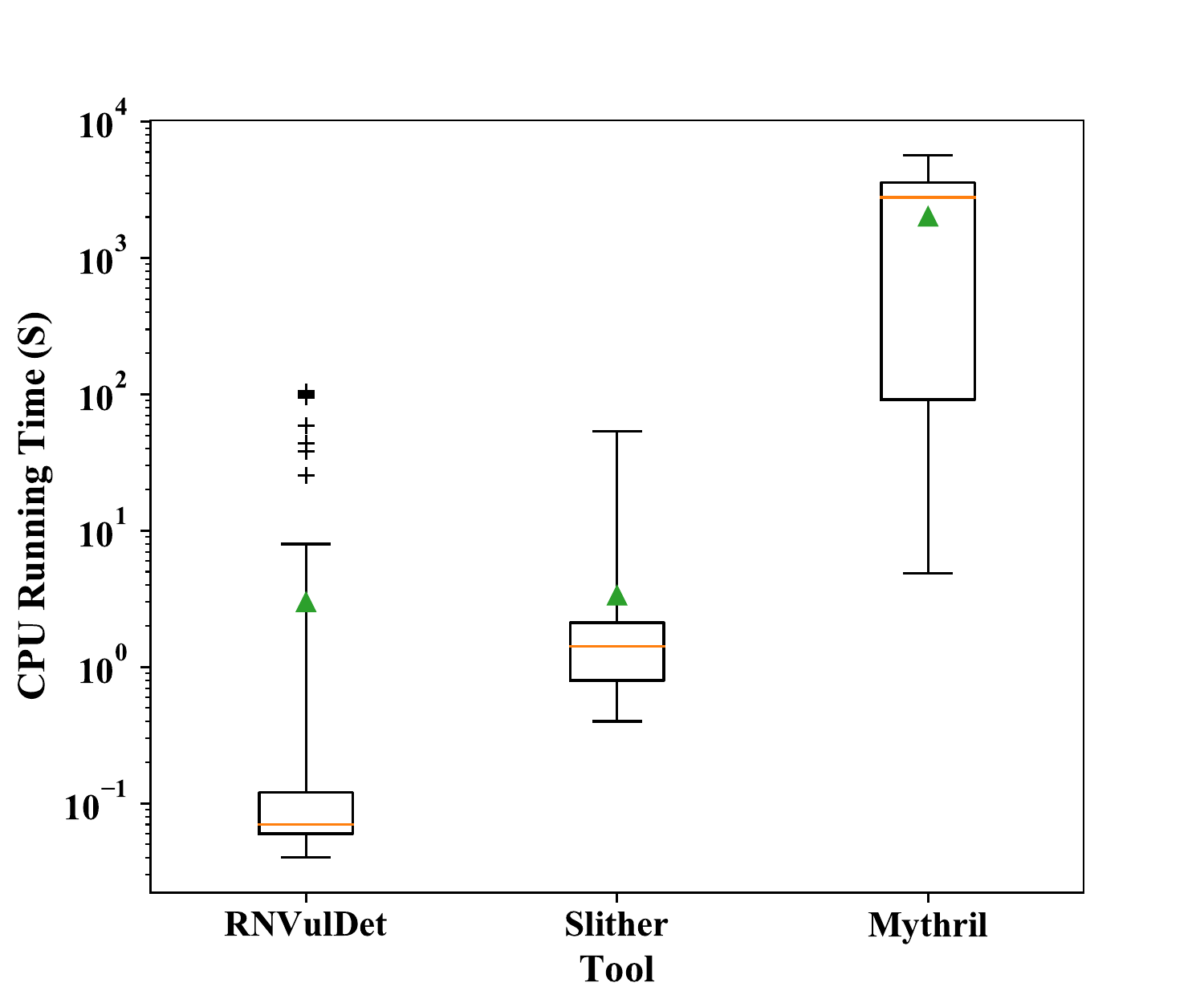}
\vspace{-0.2em}
\caption{ Boxplot visualizes the CPU running time of each tool in analyzing smart contracts. } 
\label{fig7}
\vspace{-1.2em}
\end{figure}

\renewcommand{\arraystretch}{1.3}
\begin{table*}
	\centering
	\caption{ { Experiment result of \emph{RNVulDet} in detecting random number attack transactions.} }
	\vspace{-0.9em}
	\resizebox{0.8\textwidth}{!}{
		\begin{tabular}{ccccc}
			\bottomrule 
			\textbf{Attack Type} & \textbf{Loss} & \textbf{Victim Contract} & \textbf{Attack Contract} & \textbf{Attack Transaction} \\
			\hline
			 \makecell[c]{Transaction Input Manipulation \\ Random Number Seed Manipulation \\ Direct Prediction}  & 8,928 ETH  & 145 & 243 & 42,107 \\
			 Transaction Rollback & 42 ETH  & 36  & 22 & 2,157 \\
			 \textbf{Total} & \textbf{8,970} ETH  & \textbf{181}  & \textbf{265} & \textbf{44,264} \\
			 \toprule
	\end{tabular} 
	} 
\label{table5}
\vspace{-0.8em}
\end{table*}

\renewcommand{\arraystretch}{1.15}
\begin{table*}
	\centering
	\caption{ Detection result of the top 10 smart contracts with losses caused by random number attacks. }
	\vspace{-0.9em}
	\resizebox{0.82\textwidth}{!}{
		\begin{tabular}{cccccc}
			\bottomrule 
			\textbf{\#} & \textbf{Contract Address} & \textbf{Contract Name} & \textbf{Loss} & \textbf{Result} & \textbf{Open-Sourced} \\
			\hline
			 1  & 0xdd9fd6b6f8f7ea932997992bbe67eabb3e316f3c  & Last Winner & 6,543.59 ETH & TP & {\color{DarkRed} \ding{55} } \\
			 2  & 0x47663541167ece0b96d9e5c60f9e470b2a20f598  & LD3D  & 894.46 ETH & TP & {\color{DarkRed} \ding{55} }  \\
			 3  & 0xa62142888aba8370742be823c1782d17a0389da1  & FoMo3D  & 615.27 ETH & TP & {\color{OliveGreen} \ding{51} }  \\
			 4  & 0xd1ceeeefa68a6af0a5f6046132d986066c7f9426  & Dice2Win  & 276.14 ETH & FP & {\color{OliveGreen} \ding{51} }  \\
			 5  & 0x8a883a20940870dc055f2070ac8ec847ed2d9918  & RatScam  & 96.63 ETH & TP & {\color{OliveGreen} \ding{51} }  \\
			 6  & 0x29488e24cfdaa52a0b837217926c0c0853db7962  & SuperCard  & 86.28 ETH & TP & {\color{OliveGreen} \ding{51} }  \\
			 7  & 0xab577eaed199d63af1aa8a03068d23c81fba0619  & Unnamed  & 77.65 ETH & TP &  {\color{DarkRed} \ding{55} }  \\
			 8  & 0x5fe5b7546d1628f7348b023a0393de1fc825a4fd  & Roulette  & 70.00 ETH & TP &  {\color{OliveGreen} \ding{51} }  \\
			 9  & 0x9f35334c9dc3c66347d33558b7cfe800380391b5  & Unnamed  & 43.91 ETH & TP &  {\color{DarkRed} \ding{55} }  \\
			 10  & 0xd05dc25d8dad48fb9cf242d812d8fb4a653adb95  & Unnamed  & 40.10 ETH & TP &  {\color{DarkRed} \ding{55} }  \\
			 \toprule
	\end{tabular} 
	} 
\label{table6}
\vspace{-1.2em}
\end{table*}

\subsubsection{Efficiency (RQ2)}
In this subsection, we systematically examine the efficiency of \emph{RNVulDet} and compare it against other methods. We calculate the CPU running time of each tool on 248 smart contracts ({i.e.,} the sum of \emph{Dataset\_1} and \emph{Dataset\_2}) and present the average execution time, lower quartile, median, and upper quartile, respectively. Quantitative experimental results of each tool are summarized in Table~\ref{table4}. From the table, we can observe that \emph{RNVulDet} is significantly more efficient than others. For instance, the average execution time of \emph{RNVulDet} is 2.98s, 2,006.04s faster than \emph{Mythril}. {It is worth noting that the dynamic tool \emph{ConFuzzius} will keep running until the preset timeout is reached. As we set a timeout of 3,600 seconds for each tool, the average CPU running time of \emph{ConFuzzius} is 3,600 seconds.}

Further, we visualize the results of CPU running time for each tool in Fig.~\ref{fig7}, which intuitively displays the distribution of CPU running time when analyzing different contracts. From the figure, we can learn that {(1)} The median ({i.e.,} the orange line) of the boxplot for \emph{RNVulDet} is near the bottom of the box, which indicates that \emph{RNVulDet} takes a relatively short time to analyze smart contracts in most cases. {(2)} The rectangular area of the boxplot for \emph{RNVulDet} is very small, which suggests that its execution time for different contracts is concentrated in a small range, thereby illustrating the stability of \emph{RNVulDet}.

In particular, we also observe that the boxplot of \emph{RNVulDet} contains 9 outliers ({i.e.,} the cross symbol in Fig.~\ref{fig7}), and the execution time is between 25.53 seconds and 104.92 seconds. By analyzing the detection results of \emph{RNVulDet}, we found that all these outliers are generated when detecting the contracts of \emph{FoMo3D} type. These outliers significantly drive up the average execution time of \emph{RNVulDet}  ({i.e.,} triangles in Fig.~\ref{fig7}). {Note that although the detection time for the \emph{FoMo3D} contract is relatively long, \emph{RNVulDet} is still able to detect the bad randomness vulnerability. Instead, \emph{Mythril} fails to reveal the bad randomness vulnerability in the \emph{FoMo3D} contract due to the path explosion problem. Furthermore, \emph{Slither} detects the bad randomness vulnerability by determining whether the random number seed is involved in a modulo operation, while the seed in the \emph{FoMo3D} contract does not participate in a modulo operation, thus generating false negatives.}

\subsubsection{Evaluation of Attack Detection (RQ3)}
We now present evaluation results on random number attack detection. We measure the effectiveness of \emph{RNVulDet} by counting the number of detected attack transactions. Particularly, we screen out the top 10 smart contracts with losses caused by random number attacks and visualize the relationship between losses and attack behaviors. Moreover, we calculate the period from when the contract was deployed to when a random number attack compromised it, providing overall insights on the exploitation of bad random vulnerabilities in Ethereum smart contracts.

First of all, we evaluate the performance of \emph{RNVulDet} in detecting random number attack transactions on \emph{Dataset\_3}. {Particularly, we find out potential attack transactions according to the defined attack patterns (which are defined in \S\ref{patterns_attack}).} Quantitative experimental results are summarized in Table~\ref{table5}. From the table, we can observe that bad randomness vulnerabilities have caused a loss of 8,970 ETH (which is worth around \$27 million at the time of writing) from July 2015 to May 2021. To be more specific, a total of 265 attack contracts have launched 44,264 attack transactions, of which 42,107 belong to random number manipulation or prediction attacks. Note that we group the detected attack transactions by victim contracts and calculate the amount of Ether lost for each victim contract. We then rank the victim contracts in descending order of loss amount, and manually confirm the attack transactions of the contracts with relatively significant losses (more than 40 Ether), instead of all 44,264 attack transactions.

Further, we present the detection results of the top 10 contracts with losses caused by random number attacks in Table~\ref{table6}, where we can see that losses in the 10 contracts reached 8,467.89 ETH, accounting for 94.4\% of the total losses of 181 detected victim contracts. Notably, \emph{RNVulDet} successfully identifies bad randomness vulnerabilities in 9 of these contracts. Only contract \emph{Dice2Win} was reported as a false positive. Through manual inspection, we found that \emph{Dice2Win} adopts the hash of a future block to generate pseudo-random numbers. Although block hash (i.e., \texttt{BLOCKHASH}) is a kind of \emph{vulnerable instruction}, using the \emph{future block hash} as a random seed is very difficult for attackers to manipulate or predict. However, \emph{RNVulDet} gives equal treatment to all \emph{vulnerable instructions} and thus considers the pseudo-random number generated by a future block hash as vulnerable, resulting in a false alarm.
Note that using other information of future blocks (e.g., \emph{block timestamp, block number,} and \emph{block difficulty}) to generate pseudo-random numbers still bears the inherent risk of being manipulated or predicted by miners. Only using a \emph{future block hash} for random number generation is generally considered safe. Since no one can know the content of a future block beforehand, it is almost impossible for miners to predict the pseudo-random number generated by a 256-bit future block hash.

\begin{figure}
\centering
\includegraphics[width=7.15cm]{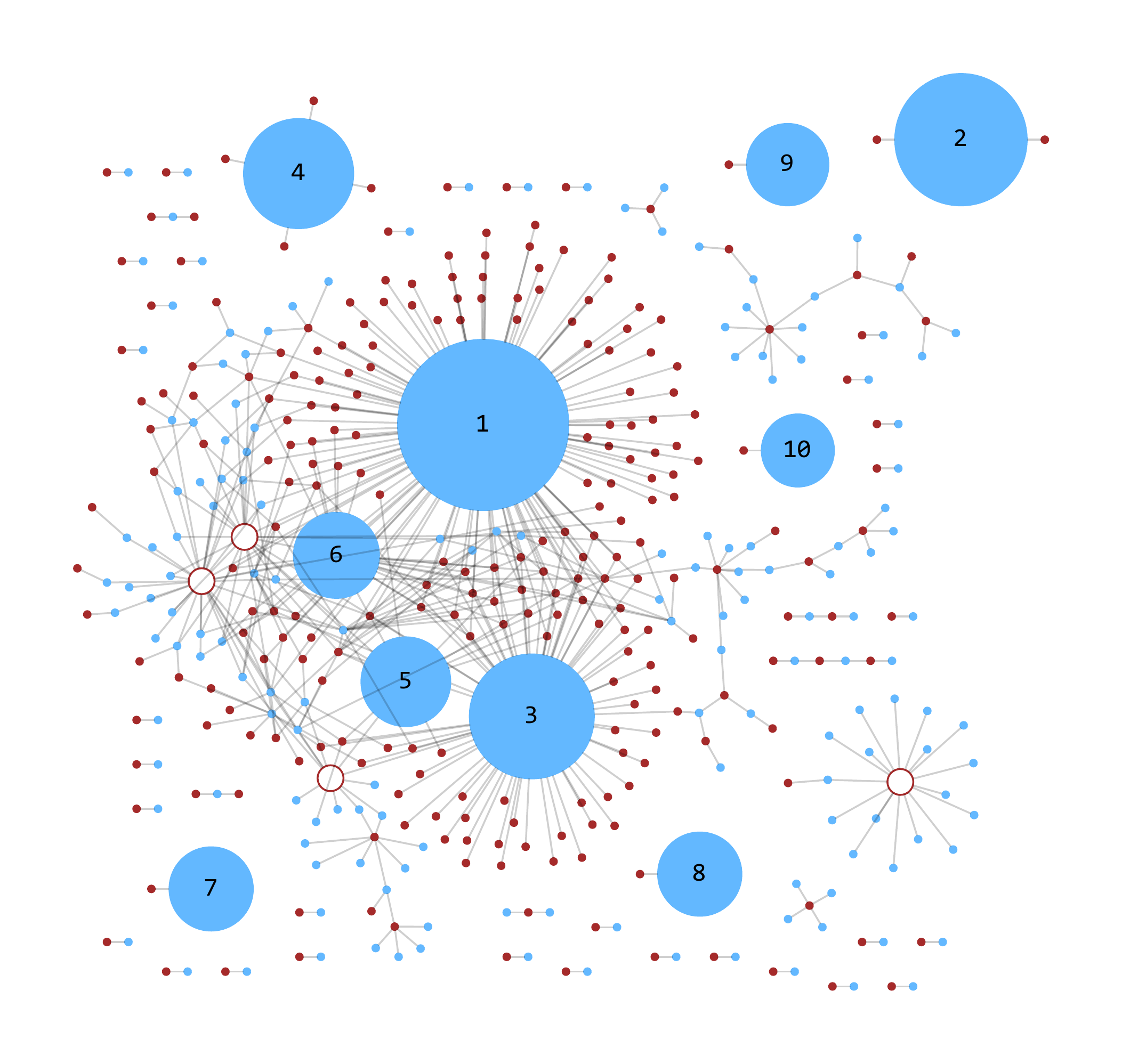}
\vspace{-0.2em}
\caption{ Visual analysis of the relationship between losses and attack behaviors.} 
\label{fig8}
\vspace{-1.2em}
\end{figure}

Fig.~\ref{fig8} further visualizes the relationship between losses and attack behaviors, where the blue and red nodes respectively represent victim contracts and attacker contracts, and the edges stand for attack behaviors. The blue nodes from 1 to 10 represent the top 10 victim contracts with the losses caused by random number attacks. The larger the area of the node, the greater the loss. The hollow red nodes denote the attackers who have compromised more than 10 contracts. From the figure, we obtain the following observations: 1) There are cases where multiple attackers attack the same victim contract, such as contracts \emph{Last Winner} and \emph{FoMo3D}, and 2) Some victim contracts, such as \emph{LD3D} and \emph{Dice2Win}, have suffered from huge losses even though they were only attacked by a few attackers.
Notably, taking Fomo3D as an example, we provide several real attack instances revealed by \emph{RNVulDet}, namely,~\cite{Intsance1,Intsance2,Intsance3}, and more detected attack transactions have been released on our Github\footnote{https://github.com/Messi-Q/RNVulDet.}.

\begin{figure}
\centering
\includegraphics[width=7.9cm]{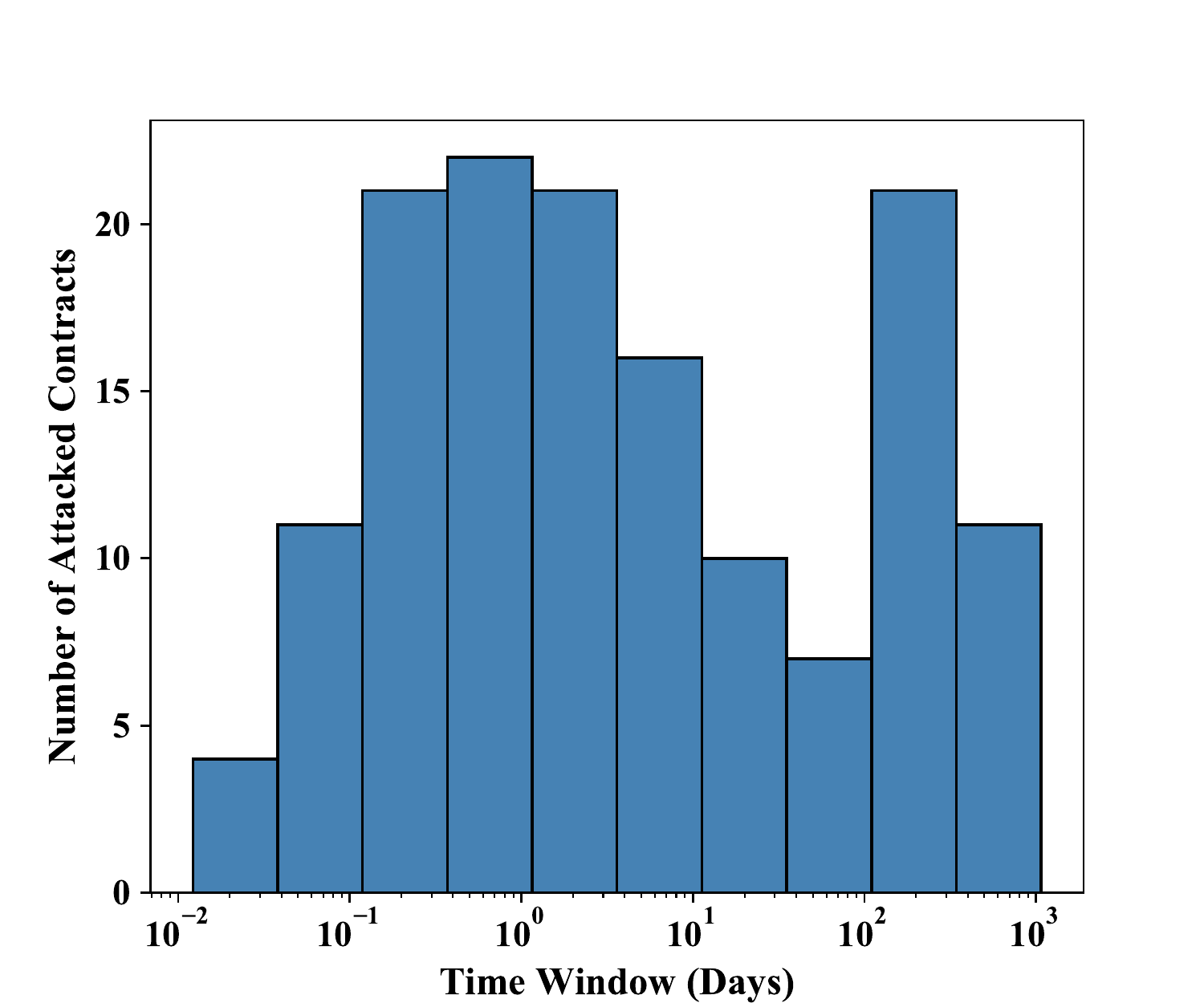}
\vspace{-0.2em}
\caption{Attack time window distribution of smart contracts with the bad randomness vulnerability.} 
\label{fig9}
\vspace{-1.2em}
\end{figure}

Finally, we depict the attack time window\footnote{ Attack time window is defined as the period from when a contract was deployed to when a random number attack compromised it. } to illustrate the overall situation of the exploitation of bad random vulnerabilities in Ethereum smart contracts. Fig.~\ref{fig9} shows the attack time window for 181 victim contracts that were compromised by random number attacks, where the shortest attack time window was 18 minutes. Put differently, it took only 18 minutes for a contract to be deployed on Ethereum until it was attacked. The longest attack time window was as high as 1,420 days. Moreover, there are 62 victim contracts whose attack time windows are less than 24 hours, which suggests the universality of random number attacks in Ethereum.

\section{Discussion}
\label{discussion}
In this section, we discuss the potential improvements of \emph{RNVulDet} as well as its limitations and completeness.

\emph{Limitations and Completeness.}\quad (1) \emph{RNVulDet} still faces a potential over-tainting problem during the taint analysis process. Since Solidity sometimes uses the exact same storage slot to store multiple variables, \emph{RNVulDet} currently fails to differentiate the different variables across the same storage slot, which may lead to over-tainting and produce false alarms. (2) Regarding the polynomial comparison of \texttt{SSTORE} and \texttt{SLOAD} instructions, the function with the same arguments may return different results because the value of some instructions, such as \texttt{PC}, \texttt{MSIZE}, and \texttt{GAS}, may change during the simulation execution. 

\emph{Future Improvement.}\quad
(1) We will consider the impact of the shared storage slot and improve the polynomial comparison in the future to reduce the false positives and negatives. (2) We will optimize the bad randomness vulnerability patterns by taking care of the future block hash and hybrid random number generation, to eliminate possible false positives. Furthermore, it is also necessary to take into account the situations such as the random number generation using simple arithmetic instructions, and the generated random number contaminating the transfer of other types of tokens, thus reducing the false negatives. (3) We will extend the techniques presented in \emph{RNVulDet}, such as Stack state examination, Memory segmentation, and Storage key-value pair comparison, to identify other types of smart contract vulnerabilities.

\section{Related Work}
\label{related_work}
Quite a few practitioners have devoted themselves to devising testing tools for smart contract vulnerability detection~\cite{oyente,perez2021smart,feist2019slither,wu2020ethscope,lu2022iquery,liu2023rethinking}. Upon scrutinizing the released implementations of existing methods, we empirically found that current vulnerability detection approaches can be roughly divided into the following three major categories.

\emph{Static Analysis.}\quad Static analysis refers to a technology of analyzing programs without actually executing them. It examines code in the absence of real input data and is able to detect potential security violations, runtime errors, and logical inconsistencies. 
Oyente~\cite{oyente} is one of the pioneer smart contract analysis tools, which uses symbolic execution to identify smart contract vulnerabilities. 
Securify~\cite{securify} performs advanced formal analysis to infer semantic facts of data flows in a smart contract and then prove the presence or absence of vulnerabilities.
Slither~\cite{feist2019slither} converts the smart contract source code into an intermediate representation of SlithIR. SlithIR uses a static single allocation form and a reduced instruction set to simplify the contract analysis process.
Defectchecker~\cite{chen2021defectchecker} is a symbolic execution-based bug checker, which extracts the CFG of smart contracts and detects vulnerabilities by analyzing instruction sequences.

\emph{Fuzzing Testing.}\quad Fuzzing has been proven to be a prominent technique for discovering software vulnerabilities in the past decades~\cite{chen2018systematic}. When applied to smart contracts, a fuzzing engine will try to generate initial seeds to form executable transactions. The feedback of fuzzing results will dynamically guide the generation of new test cases. This process repeats until a stopping criterion is satisfied. Eventually, the fuzzer will analyze the results generated during fuzzing and report to users. 
Contractfuzzer~\cite{jiang2018contractfuzzer} is the first to apply fuzzing techniques to smart contracts. It identifies vulnerabilities by monitoring runtime behaviors during fuzzing.
ILF~\cite{he2019learning} proposes a new approach for learning an effective yet fast fuzzer from symbolic execution by phrasing the learning task in the framework of imitation learning.
Harvey~\cite{wustholz2020harvey} extends standard greybox fuzzing with a method for predicting new inputs that are more likely to cover new paths or reveal vulnerabilities in smart contracts.
sFuzz~\cite{nguyen2020sfuzz} presents an efficient and lightweight multi-objective adaptive strategy to dig out potential vulnerabilities hidden in those hard-to-cover branches.

\emph{Machine Learning.}\quad Recent years have witnessed an increasing practice of detecting program security vulnerabilities using machine learning techniques~\cite{xing2020new,wu2022towards}. The advancement of machine learning has promoted the emergence of multiple vulnerability detection methods.
Contractward~\cite{wang2020contractward} extracts bigram features from the smart contract opcode and adopts a variety of machine learning algorithms to detect bugs in smart contracts.
$S$-gram~\cite{liu2018s} introduces a novel semantic-aware security auditing technique for analyzing smart contracts. The key insight behind $S$-gram is a combination of $N$-gram language modeling and lightweight static contract analysis.
Liu \emph{et al.}~\cite{liu2021combining} propose to cast the source code of a smart contract into a contract graph and construct a temporal-message-propagation graph neural network to identify vulnerabilities.
Xue \emph{et al.}~\cite{xue2022xfuzz} present a machine learning-guided smart contract fuzzing framework, which leverages machine learning predictions to guide fuzzers for vulnerability detection.

\section{Conclusion}
\label{conclusion}
In this work, we present a special focus on random numbers in Ethereum smart contracts. We investigate the principles behind different pseudo-random number generation strategies and organize them into a taxonomy. We propose \emph{RNVulDet}, a tool that can automatically identify bad randomness vulnerabilities and detect random number attack transactions, with a very low false positive and negative rate. Specifically, \emph{RNVulDet} consists of four key components, {i.e.,} stack state examination, memory segmentation, storage key-value pair comparison, and transaction replay. An analysis of the real-world transactions in Ethereum revealed that 265 attackers have already issued 44,264 random number attack transactions, which caused more than \$27 million in losses. It is worth mentioning that these numbers solely provide a lower bound and thus might only reflect the tip of the iceberg. Experimental results on three kinds of datasets demonstrate that \emph{RNVulDet} significantly surpasses state-of-the-art vulnerability detection tools by a large margin. Our implementation and dataset are released to facilitate future research. The presented designs in \emph{RNVulDet} can also be transferable to detect other types of vulnerabilities in Ethereum smart contracts.

\section*{Acknowledgment}
This work was supported by the Key R\&D Program of Zhejiang Province (No. 2022C01086), and the National Natural Science Foundation of China (No.62172360, No.U21A20467).

\footnotesize
\bibliographystyle{IEEEtran}
\bibliography{bare_jrnl_compsoc}

\end{sloppypar}

\end{document}